\documentclass[letterpaper, english, aps, prl, reprint, twocolumn, superscriptaddress]{revtex4-2}
\usepackage[T1]{fontenc}
\usepackage{color}
\usepackage{amsmath}
\usepackage{amssymb}
\usepackage{graphicx}
\usepackage{multirow}
\usepackage{booktabs}
\usepackage{capt-of}
\usepackage[latin9]{inputenc}
\usepackage{babel}
\usepackage{calc}
\usepackage{float}
\usepackage{dcolumn}

\makeatletter
\usepackage[colorlinks=true, letterpaper=true, pdfstartview=FitV, linkcolor=blue, citecolor=blue, urlcolor=blue]{hyperref}
\renewcommand{\fnum@figure}{FIG. \thefigure}
\renewcommand{\fnum@table}{TABLE. \thetable}
\makeatother

\begin{document}

\title{Quantum metric-based optical selection rules}

\author{Yongpan Li}
\affiliation{Centre for Quantum Physics, Key Laboratory of Advanced Optoelectronic Quantum Architecture and Measurement (MOE), School of Physics, Beijing Institute of Technology, Beijing, 100081, China}
 
\author{Cheng-Cheng Liu}
\email{ccliu@bit.edu.cn}
\affiliation{Centre for Quantum Physics, Key Laboratory of Advanced Optoelectronic Quantum Architecture and Measurement (MOE), School of Physics, Beijing Institute of Technology, Beijing, 100081, China}

\begin{abstract}
The optical selection rules dictate symmetry-allowed/forbidden transitions, playing a decisive role in engineering exciton quantum states and designing optoelectronic devices. While both the real (quantum metric) and imaginary (Berry curvature) parts of quantum geometry contribute to optical transitions, the conventional theory of optical selection rules in solids incorporates only Berry curvature. Here, we propose quantum metric-based optical selection rules. We unveil a universal quantum metric-oscillator strength correspondence for linear polarization of light and establish valley-contrasted optical selection rules that lock orthogonal linear polarizations to distinct valleys. Tight-binding and first-principles calculations confirm our theory in two models (altermagnet and Kane-Mele) and monolayer $d$-wave altermagnet $\mathrm{V_2SeSO}$. This work provides a quantum metric paradigm for valley-based spintronic and optoelectronic applications.
\end{abstract}

\maketitle

\begin{figure}[t]
\centering
\includegraphics[width=1.0\linewidth]{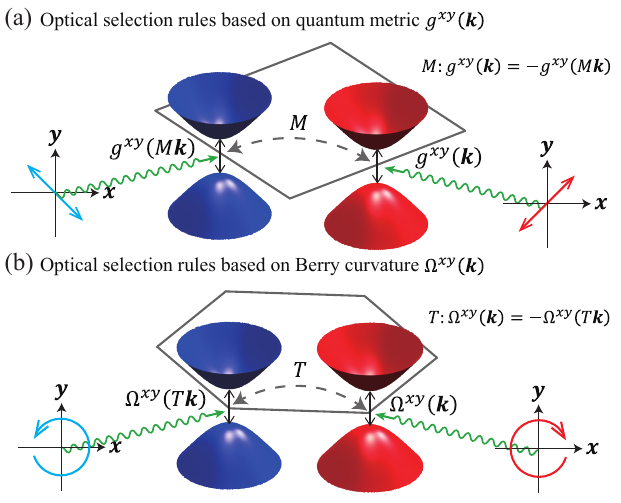}
\caption{Schematics of the quantum metric-based optical selection rules for linearly polarized light (a) and the Berry curvature-based optical selection rules for circularly polarized light (b). (a) The two valleys are located at mirror-invariant lines or rotation axes of in-plane two-fold rotational symmetry and related by $M$ symmetry (e.g., mirror symmetry or four-fold rotational symmetry) with non-zero quantum metric $g^{xy}(\boldsymbol{k})=-g^{xy}(M\boldsymbol{k})$ and vanishing Berry curvature. For two orthogonal linearly polarized lights along specific directions, one (another) light exclusively excites electrons at one (the other) valley. (b) The two valleys are located at $n$-fold rotation axes ($n\geq3$) and related by $T$ symmetry (e.g., time-reversal symmetry or mirror symmetry) with non-zero Berry curvature $\Omega^{xy}(\boldsymbol{k})=-\Omega^{xy}(T\boldsymbol{k})$ and vanishing quantum metric. The left (right) circularly polarized light exclusively excites electrons at one (the other) valley.}
\label{fig:schematic}
\end{figure}

\textit{Introduction.}---In atoms, optical selection rules govern allowed transitions based on atomic orbitals' orbital angular momentum and symmetries~\cite{hasegawa1961optical}. Inspired by the close connection between optical selection rules and orbital angular momentum in atoms, the discovery that orbital angular momentum in solids is closely related to Berry curvature led to the emergence of Berry curvature-based optical selection rules for circularly polarized light~\cite{PhysRevLett.95.137205,PhysRevLett.95.137204,PhysRevB.74.024408,yao_valleydependent_2008, cao_valleyselective_2012, xiao_coupled_2012}, as schematically illustrated in Fig.~\ref{fig:schematic}(b). These rules are well established and widely studied, particularly in graphene and transition metal dichalcogenides~\cite{yao_valleydependent_2008, cao_valleyselective_2012, xiao_coupled_2012}. 
However, both the Berry curvature and the quantum metric---representing the imaginary and real parts of quantum geometry, respectively---have been shown to qualitatively influence optical transitions~\cite{provost_riemannian_1980,shapere1989geometric,ma_abelian_2010,aversa_nonlinear_1995, de2017quantized, ahn_riemannian_2022, verma2024instantaneous, ezawa2024analytic, komissarov_quantum_2024}. Consequently, the quantum metric is also expected to give rise to optical selection rules. Moreover, as both linear and circular polarization are fundamental and equally significant polarization states of light, optical selection rules should extend beyond circularly polarized light to include linearly polarized light, considering that some works~\cite{PhysRevX.12.021055,PhysRevB.93.045431,PhysRevB.94.035304,chen2018valley,PhysRevLett.124.037701,bzzy-ngcs} have noted certain valleys can exhibit complete linear polarization selection.

In this work, we establish quantum metric-based optical selection rules for linearly polarized light, linking the quantum metric directly to oscillator strength for linear polarization of light, as schematically illustrated in Fig.~\ref{fig:schematic}(a). For a valley with mirror or mirror-like symmetry, the electrons in the valley couple exclusively to linear polarization aligned parallel/perpendicular to its mirror plane. When two valleys exhibit mutually perpendicular mirror planes and are connected by symmetry like mirror symmetry and four-fold rotational symmetry, we demonstrate valley-contrasted selection rules: orthogonal linear polarizations selectively excite distinct valleys. We validate our theory in an altermagnet (AM) model and the Kane-Mele (KM) model, and further probe its realization in monolayer $d$-wave altermagnet $\mathrm{V_2SeSO}$ via density functional theory (DFT). This work extends the geometric origin of optical selection rules beyond the Berry curvature paradigm.

\textit{General theory.}---Here, we derive the quantum metric-based optical selection rules for linearly polarized light. The matrix elements of the quantum metric tensor $\boldsymbol{g}(\boldsymbol{k})$ can be given by~\cite{provost_riemannian_1980,shapere1989geometric,ma_abelian_2010,aversa_nonlinear_1995}
\begin{equation}\label{eq:qm}
g^{ab}_{nm}(\boldsymbol{k})\equiv \dfrac{1}{2}\dfrac{v^a_{nm,\boldsymbol{k}}v^b_{mn,\boldsymbol{k}}+v^b_{nm,\boldsymbol{k}}v^a_{mn,\boldsymbol{k}}}{\omega_{nm,\boldsymbol{k}}^2},
\end{equation}
where $a$, $b$, and $c$ label Cartesian directions, and $v^a$ is the matrix representation of the velocity operator in the $a$ direction.

Without loss of generality, one can assume the light propagates along the $z$-direction. The matrix elements of the quantum metric tensor, i.e., $g^{xy}_{nm}(\boldsymbol{k})$, can be rewritten in terms of the oscillator strength for the transition between states $n$ and $m$ as (Supplementary Materials~\cite{SM} Sec. I )
\begin{equation}
g^{xy}_{nm}(\boldsymbol{k})=\dfrac{\hbar(f_{nm}^{(\hat{\boldsymbol{x}}+\hat{\boldsymbol{y}})}(\boldsymbol{k})-f_{nm}^{(\hat{\boldsymbol{x}}-\hat{\boldsymbol{y}})}(\boldsymbol{k}))}{8m_e\omega_{nm,\boldsymbol{k}}},
\end{equation}
where $m_e$ is the electron mass and $\hbar\omega_{nm,\boldsymbol{k}}\equiv \hbar\omega_{n,\boldsymbol{k}}-\hbar\omega_{m,\boldsymbol{k}}$ is the energy difference between the $m$th and $n$th bands. $f_{nm}^{(\hat{\boldsymbol{x}}+\hat{\boldsymbol{y}})}(\boldsymbol{k})$ and $f_{nm}^{(\hat{\boldsymbol{x}}-\hat{\boldsymbol{y}})}(\boldsymbol{k})$ are the $k$-resolved oscillator strengths, i.e., the optical absorption strengths, for linear polarization along the $\hat{\boldsymbol{x}}+\hat{\boldsymbol{y}}$ and $\hat{\boldsymbol{x}}-\hat{\boldsymbol{y}}$, respectively, and read~\cite{yu_fundamentals_2010a, sakurai2020modern,souza_dichroic_2008}
\begin{equation}
f_{nm}^{(\hat{\boldsymbol{\epsilon}})}=\dfrac{2m_e\left|\langle n|\hat{\boldsymbol{\epsilon}}\cdot\boldsymbol{v}_{\boldsymbol{k}}|m\rangle\right|^2}{\hbar\omega_{nm,\boldsymbol{k}}}.
\end{equation}
Here $\hat{\boldsymbol{\epsilon}}\cdot\boldsymbol{v}_{\boldsymbol{k}}=v^x_{\boldsymbol{k}}+v^y_{\boldsymbol{k}}$ or $v^x_{\boldsymbol{k}}-v^y_{\boldsymbol{k}}$.

At frequency $\omega$, the $k$-resolved degree of linear polarization in the $\hat{\boldsymbol{x}}+\hat{\boldsymbol{y}}$ and $\hat{\boldsymbol{x}}-\hat{\boldsymbol{y}}$ directions is given by
\begin{eqnarray}\label{eq:eta}
\eta(\boldsymbol{k})&\equiv&\dfrac{f_{\mathrm{total}}^{(\hat{\boldsymbol{x}}+\hat{\boldsymbol{y}})}(\boldsymbol{k})-f_{\mathrm{total}}^{(\hat{\boldsymbol{x}}-\hat{\boldsymbol{y}})}(\boldsymbol{k})}{f_{\mathrm{total}}^{(\hat{\boldsymbol{x}}+\hat{\boldsymbol{y}})}(\boldsymbol{k})+f_{\mathrm{total}}^{(\hat{\boldsymbol{x}}-\hat{\boldsymbol{y}})}(\boldsymbol{k})}\nonumber\\
&=&\dfrac{2\sum_{nm}g^{xy}_{nm}(\boldsymbol{k})\delta(\omega-\omega_{mn,\boldsymbol{k}})}{\sum_{nm}\left(g^{xx}_{nm}(\boldsymbol{k})+g^{yy}_{nm}(\boldsymbol{k})\right)\delta(\omega-\omega_{mn,\boldsymbol{k}})},
\end{eqnarray}
where $f_{\mathrm{total}}^{(\hat{\boldsymbol{\epsilon}})}=\sum_{nm}f_{nm}^{(\hat{\boldsymbol{\epsilon}})}\delta(\omega-\omega_{mn,\boldsymbol{k}})$ and the summation $\sum_{nm}$ implicitly means $n$ runs over occupied states and $m$ over unoccupied states.

The connections between the oscillator strengths and the quantum metric tensor provide a mechanism to engineer the degree of linear polarization $\eta(\boldsymbol{k})$ in given bands through symmetry. Here we focus on two classes of symmetries: 1. Symmetries connecting distinct $k$-points $\boldsymbol{k}$ and $\boldsymbol{k}'$, for which $\eta(\boldsymbol{k}) = -\eta(\boldsymbol{k}')$. 2. Symmetries leaving $\boldsymbol{k}$ invariant, for which $\eta(\boldsymbol{k})=\pm1$.

In the presence of the mirror symmetry $M$ ($M=M_x,M_y$), $g^{xy}_{nm}(\boldsymbol{k})$ acquires a counterpart at $M\boldsymbol{k}$ with an opposite value, i.e., $M:g^{xy}_{nm}(\boldsymbol{k})=-g^{xy}_{nm}(M\boldsymbol{k})$. Combined with $g^{xx}(\boldsymbol{k})=g^{xx}(M\boldsymbol{k})$ and $g^{yy}(\boldsymbol{k})=g^{yy}(M\boldsymbol{k})$, Eq.~(\ref{eq:eta}) dictates $\eta(\boldsymbol{k})=-\eta(M\boldsymbol{k})$. Therefore, the two valleys connected by mirror symmetry $M$ exhibit opposite degrees of linear polarization.

\begin{figure*}[t]
\centering
\includegraphics[width=1.0\linewidth]{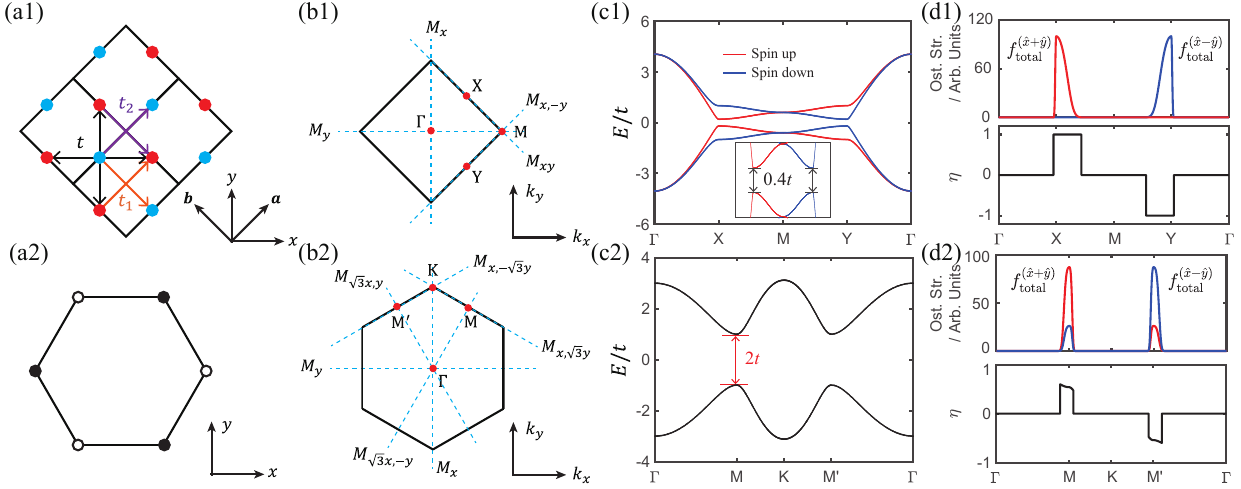}
\caption{The quantum metric-based optical selection rule for linearly polarized light in the altermagnet model and Kane-Mele model. (a) Schematic of the altermagnet model (a1) and Kane-Mele model (a2). The red dots and blue dots denote the spin-up sublattice and spin-down sublattice, respectively. (b) Relevant mirror symmetries in the Brillouin zone for the altermagnet model (b1) and Kane-Mele model (b2). (c) The corresponding energy bands. The insert in (c1) shows the magnified view of the two valleys.  (d) The corresponding $k$-resolved oscillator strengths and degree of linear polarization $\eta(\boldsymbol{k})$. Parameters: The photon energies are $0.4t$ for the altermagnet model and $2t$ for the Kane-Mele model, while the smearing parameters are $0.1t$ for both. $t_1=-t_2=0.1t$, $\lambda_{\mathrm{Z}}=0.6t$, and $t_{\mathrm{SOC}}=0.6t$.}
\label{fig:tbmodel}
\end{figure*}

We now consider the scenario of optical transitions involving two energy levels, labeled as $E_c$ and $E_v$, that are non-degenerate. $\eta(\boldsymbol{k})$ can be simplified to
\begin{equation}\label{eq:dop_twoband}
\eta(\boldsymbol{k})=\dfrac{2g^{xy}_{vc}(\boldsymbol{k})}{g^{xx}_{vc}(\boldsymbol{k})+g^{yy}_{vc}(\boldsymbol{k})}.
\end{equation}
The quantum metric tensor is positive semi-definite~\cite{provost_riemannian_1980,shapere1989geometric, peotta_superfluidity_2015,PhysRevB.90.165139,PhysRevB.104.045103}, i.e., $g_{vc}^{xx}(\boldsymbol{k})g_{vc}^{yy}(\boldsymbol{k})-g_{vc}^{xy}(\boldsymbol{k})g_{vc}^{yx}(\boldsymbol{k})\ge0$, and is symmetric under swapping of $x$ and $y$, i.e., $g_{vc}^{xy}(\boldsymbol{k})=g_{vc}^{yx}(\boldsymbol{k})$. Therefore, $-1\le\eta(\boldsymbol{k})\le1$ and we find that $\eta(\boldsymbol{k})$ takes the values of $\pm1$ when $\boldsymbol{k}$ lies on the mirror-invariant planes, with $g^{xy}_{vc}(\boldsymbol{k})=g^{xx}_{vc}(\boldsymbol{k})=g^{yy}_{vc}(\boldsymbol{k})$ or $-g^{xy}_{vc}(\boldsymbol{k})=g^{xx}_{vc}(\boldsymbol{k})=g^{yy}_{vc}(\boldsymbol{k})$. Specifically,  the mirror symmetry $M_{xy}:(x,y)\to(-y,-x)$ imposes $\langle m|v^x_{\boldsymbol{k}}|n\rangle=\langle M_{xy}m|(-v^y_{\boldsymbol{k}})|M_{xy}n\rangle=-e^{i(\theta_{n}-\theta_{m})}\langle m|v^y_{\boldsymbol{k}}|n\rangle$, where $e^{i\theta_n}$ is the mirror eigenvalue for the state $|n\rangle$, and thus
\begin{equation}\label{eq:mxy}
M_{xy}:-e^{i(\theta_{v}-\theta_{c})}g^{xy}_{vc}(\boldsymbol{k})=g^{xx}_{vc}(\boldsymbol{k})=g^{yy}_{vc}(\boldsymbol{k}).
\end{equation}
For spinful systems, $\theta = \pm\pi/2$, while for spinless systems, $\theta = 0\ \mathrm{or}\ \pi$. Similarly, the mirror symmetry $M_{x,-y}:(x,y)\to(y,x)$ enforces
\begin{equation}\label{eq:mx-y}
M_{x,-y}:e^{i(\theta_{v}-\theta_{c})}g^{xy}_{vc}(\boldsymbol{k})=g^{xx}_{vc}(\boldsymbol{k})=g^{yy}_{vc}(\boldsymbol{k}).
\end{equation}
Substituting Eq.~(\ref{eq:mxy}) and Eq.~(\ref{eq:mx-y}) separately into Eq.~(\ref{eq:dop_twoband}), we find that both $M_{xy}$ and $M_{x,-y}$ can ensure $\eta = \pm1$. Therefore, the specific mirror symmetry $M$ $(M=M_{xy},M_{x,-y})$ impose a selection rule that the linearly polarized light polarized in $\hat{\boldsymbol{x}}+\hat{\boldsymbol{y}}$ or $\hat{\boldsymbol{x}}-\hat{\boldsymbol{y}}$ is exclusively allowed in optical transitions at the mirror invariant plane of $M$, i.e., $\boldsymbol{k}=M\boldsymbol{k}$. As a result, a valley exhibits complete linear polarization selectivity if the optical transition at the valley involves nondegenerate two energy levels and the valley has mirror symmetry and nonzero quantum metric (SM \cite{SM} Secs. I and II). Note that if $\boldsymbol{k}$ simultaneously lies in the mirror invariant planes of $M_{xy}$ and $M_{x,-y}$, the complete linear polarization selectivity can still emerge when the mirror eigenvalues of the conduction and valence bands are identical for one mirror symmetry but opposite for the other, as derived from Eqs. (\ref{eq:mxy}) and (\ref{eq:mx-y}). 

Here we revisit the Berry curvature-based optical selection rules for circularly polarized light with the details in SM~\cite{SM} Sec. I. There is a link between the Berry curvature $\Omega_{nm}^{xy}$ and the oscillator strengths for left- and right-circularly polarized light $f_{nm}^{(\hat{\boldsymbol{x}}\pm i\hat{\boldsymbol{y}})}(\boldsymbol{k})$, i.e., $\Omega^{xy}_{nm}(\boldsymbol{k})=-\hbar(f_{nm}^{(\hat{\boldsymbol{x}}+i\hat{\boldsymbol{y}})}(\boldsymbol{k})-f_{nm}^{(\hat{\boldsymbol{x}}-i\hat{\boldsymbol{y}})}(\boldsymbol{k}))/{4m_e\omega_{nm,\boldsymbol{k}}}$. The degree of circular polarization $\eta_{nm}(\boldsymbol{k})$ for the optical transition between states $n$ and $m$ is given by $\eta_{nm}(\boldsymbol{k})=-\Omega^{xy}_{nm}(\boldsymbol{k})/(g^{xx}_{nm}(\boldsymbol{k})+g^{yy}_{nm}(\boldsymbol{k}))$. A complete circular polarization selectivity emerges with $\eta(\boldsymbol{k})=\pm1$ at a valley with $n$-fold ($n\geq3$) rotational symmetry, non-zero Berry curvature, and nondegenerate two energy levels (SM~\cite{SM} Sec. I D). The left or right circularly polarized light couples exclusively to the optical transitions at a certain valley. When two such valleys are connected by time-reversal symmetry or mirror symmetry, there are the valley-contrasted optical selection rules, i.e., one (the other) valley selectively couples with left (right) circularly polarized light, as shown in Fig.~\ref{fig:schematic}(b).

The complete linear polarization selectivity cannot be exhibited in valleys where the optical transitions involve two energy levels that are degenerate. The reason is that the inequality arising from the positive semidefiniteness of the quantum metric tensor holds with equality only when the optical transitions involve nondegenerate two energy levels~\cite{peotta_superfluidity_2015,PhysRevB.90.165139,PhysRevB.104.045103}, and thus $\eta(\boldsymbol{k})$ cannot be $\pm1$. The complete circular polarization selectivity in Berry curvature-based optical selection rules also can not be achieved in valleys with degenerate two energy levels due to the positive semidefiniteness of the quantum geometry tensor (SM~\cite{SM} Sec. I D). However, optical transitions of a specific frequency generally involve nondegenerate two energy levels, so complete polarization selectivity can often be achieved.

The mirror symmetry in the above derivations can be extended to other symmetries, which impose similar symmetry constraints. We collectively refer to these symmetries as mirror-like symmetries. For example, the mirror symmetry $M_x$ can be replaced by the glide symmetry $M_x\tau$ and the two-fold rotational symmetry $C_{2y}$.

\textit{Tight-binding models.}---Here we show the quantum metric-based linear polarization optical selection rules in an AM model and the KM model.

We first consider a tight-binding model of the $d$-wave AM as shown in Fig.~\ref{fig:tbmodel}(a1). The Hamiltonian is given by
\begin{eqnarray}\label{Eq:model}
H_{\mathrm{AM}}&=&t\sum_{\langle i,j\rangle} (c_{1,i}^\dag c_{2,j}+\mathrm{h.c.})+\lambda_{\mathrm{Z}}\sum_{\alpha,i} (-1)^\alpha c_{\alpha,i}^\dag s_z c_{\alpha,i}\nonumber\\
&&+t_1\sum_{i}(c_{1,i}^\dag c_{1,i+\boldsymbol{a}}+c_{2,i}^\dag c_{2,i+\boldsymbol{b}}+\mathrm{h.c.})\nonumber\\
&&+t_2\sum_{i}(c_{2,i}^\dag c_{2,i+\boldsymbol{a}}+c_{1,i}^\dag c_{1,i+\boldsymbol{b}}+\mathrm{h.c.}).
\end{eqnarray}
$c_{\alpha,i}^\dag$ with $\alpha=1,2$ creates electrons at sublattice $\alpha$ of the unit cell $i$. The first term is the nearest neighbor hopping between different sublattices. The second term is the staggered Zeeman field mimicking the N\'eel order, and $s_z$ is the spin Pauli matrix. The third and fourth terms are the next nearest neighbor hopping between the same sublattices, and $\boldsymbol{a}$ and $\boldsymbol{b}$ are the lattice vectors [See Fig.~\ref{fig:tbmodel}(a1)]. Because of the absence of the spin-orbit coupling (SOC), the AM model is described by the spin group symmetry~\cite{smejkal_conventional_2022,liu_spingroup_2022}. The sublattices with opposite magnetization are connected through a $C_4$ rotation. This model features two valleys at X and Y and exhibits spin-valley locking as shown in Fig.~\ref{fig:tbmodel}(c1). There are four relevant mirror symmetries, i.e., $M_x$, $M_y$, $M_{xy}$, and $M_{x,-y}$, which are shown in Fig.~\ref{fig:tbmodel}(b1).

To first order in $k$, the two-band $k\cdot p$ Hamiltonians at X and Y take the forms
\begin{equation}
H_{\mathrm{AM,X/Y}}=-\sqrt{2}t(k_x+\xi k_y)\sigma_x+\xi(\lambda_Z-2t_1+2t_2)\sigma_z,
\end{equation}
where $\xi=1$ at X and $\xi=-1$ at Y. The velocity operators are given by $v^x=\xi v^y=-\sqrt{2}t\sigma_x/\hbar$. According to  Eqs.~(\ref{eq:qm})(\ref{eq:dop_twoband}), we have $g^{xy}_{vc}=g^{xx}_{vc}=g^{yy}_{vc}$ at X and $-g^{xy}_{vc}=g^{xx}_{vc}=g^{yy}_{vc}$ at Y, and thus $\eta(\mathrm{X})=-\eta(\mathrm{Y})=1$. Therefore, optical transitions at the X valley exclusively admit $\hat{\boldsymbol{x}}+\hat{\boldsymbol{y}}$-polarized light, while the Y valley selectively responds to $\hat{\boldsymbol{x}}-\hat{\boldsymbol{y}}$ polarization. The numerically calculated oscillator strengths and degree of polarization are shown in Fig.~\ref{fig:tbmodel}(d1), and $\eta(\boldsymbol{k})$ are exactly $\pm1$ on the mirror-invariant lines.

Then we consider the KM model as shown in Fig.~\ref{fig:tbmodel}(a2). The Hamiltonian is given by~\cite{PhysRevLett.95.146802,PhysRevLett.95.226801}
\begin{equation}
H_{\mathrm{KM}}=-t\sum_{\langle ij\rangle}c^\dag_{i}c_{j}+it_{\mathrm{SOC}}\sum_{\langle\!\langle ij\rangle\!\rangle}v_{ij}c_{i}^\dag s_zc_{j}.
\end{equation}
$c_{i}^\dag$ creates electrons at site $i$. The first term is the nearest neighbor hopping. The second term is the effective SOC term, which shifts the valley position from K to M when $t_{\mathrm{SOC}}> \sqrt{3}t/9$. $\nu_{ij}=(\boldsymbol{d}_i\times\boldsymbol{d}_j/|\boldsymbol{d}_i\times\boldsymbol{d}_j|)_z=\pm1$, where $\boldsymbol{d}_i$ and $\boldsymbol{d}_j$ are two nearest bonds connecting the next-nearest neighbors $i$ and $j$. The two-fold band degeneracy persists throughout the Brillouin zone due to the space-time inversion symmetry. The valleys of the KM model emerge at M and M$'$ as shown in Fig.~\ref{fig:tbmodel}(c2). The relevant mirror symmetries are $M_x$, $M_y$, $M_{x,\sqrt{3}y}$, $M_{\sqrt{3}x,-y}$, $M_{x,-\sqrt{3}y}$, and $M_{\sqrt{3}x,y}$ which are shown in Fig.~\ref{fig:tbmodel}(b2).

The four-band $k\cdot p$ Hamiltonians at M and M$'$ with the first order in $k$ read
\begin{eqnarray}
H_{\mathrm{KM,M/M'}}&=&-\frac{t}{2}(\sigma_xs_0+\sqrt{3}\xi\sigma_ys_0)\nonumber\\
&&+\frac{\sqrt{3}t}{6}(\xi k_x+\sqrt{3} k_y)(\sqrt{3}\sigma_xs_0-\xi\sigma_ys_0)\nonumber\\
&&-2t_{\mathrm{SOC}}(\sqrt{3}\xi k_x-k_y)\sigma_zs_z,
\end{eqnarray}
where $\xi=1$ at M and $\xi=-1$ at M$'$. The velocity operators read $v^x=\xi t\sigma_xs_0/2\hbar-\sqrt{3}t\sigma_ys_0/6\hbar-2\sqrt{3}\xi t_{\mathrm{SOC}}\sigma_{z}s_z/\hbar$ and $v^y=\sqrt{3}t\sigma_xs_0/2\hbar-\xi t\sigma_ys_0/2\hbar+2t_{\mathrm{SOC}}\sigma_{z}s_z/\hbar$. The degrees of linear polarization at M and M$'$ are given by $\eta(\mathrm{M})=-\eta(\mathrm{M}')=\sqrt{3}(t^2-12t^2_{\mathrm{SOC}})/2(t^2+12t^2_{\mathrm{SOC}})$. The numerically calculated oscillator strengths and degree of polarization are shown in Fig.~\ref{fig:tbmodel}(d2).

For linearly polarized light with $\hat{\boldsymbol{x}}+\sqrt{3}\hat{\boldsymbol{y}}$ and $\sqrt{3}\hat{\boldsymbol{x}}-\hat{\boldsymbol{y}}$ polarizations at M, the degree of polarization is given by $\eta(\mathrm{M})=(t^2-12t^2_{\mathrm{SOC}})/(t^2+12t^2_{\mathrm{SOC}})$. Similarly, the degree of polarization for linearly polarized light with $\hat{\boldsymbol{x}}-\sqrt{3}\hat{\boldsymbol{y}}$ and $\sqrt{3}\hat{\boldsymbol{x}}+\hat{\boldsymbol{y}}$ at M$'$ are given by $\eta(\mathrm{M}')=-(t^2-12t^2_{\mathrm{SOC}})/(t^2+12t^2_{\mathrm{SOC}})$. $\eta(\mathrm{M})$ ($\eta(\mathrm{M'})$) can take the value of 1 (-1) when $t_{\mathrm{SOC}}=0$. See SM~\cite{SM} Sec. III for more details.

\begin{figure}[t]
\centering
\includegraphics[width=1.0\linewidth]{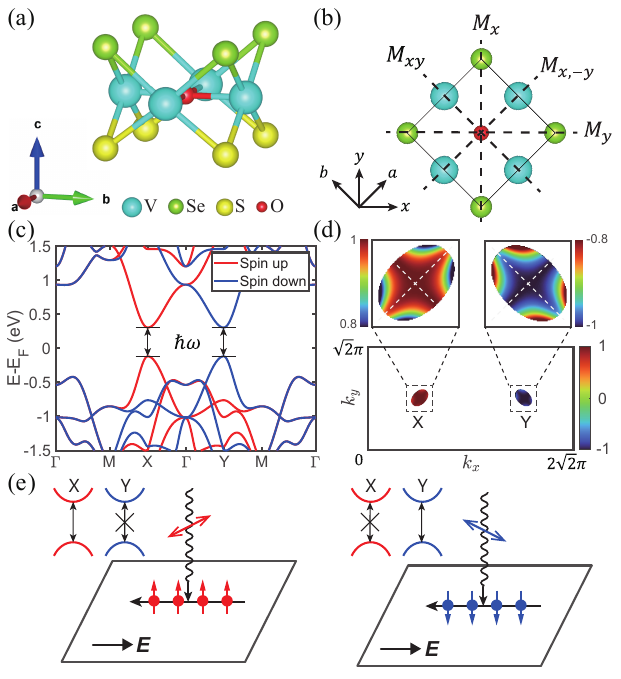}
\caption{Valley-contrasted quantum metric-based optical selection rules and fully spin-polarized currents induced by linearly polarized light in altermagnet $\mathrm{V_2SeSO}$. (a) The crystal structure. (b) The mirror symmetries. (c) The spin-splitting band structures. (d) The $k$-resolved degree of linear polarization $\eta(\boldsymbol{k})$ for photon energy $\hbar\omega$ (shown in (c)) with a smearing parameter of 0.05 eV. The insets are $\eta(\boldsymbol{k})$ around X and Y, with the white dashed lines the mirror invariant lines. (e) Schematic of the generation of the fully spin-polarized current based on valley-contrasted quantum metric-based complete optical selection. Here only the electrons are drawn, while the holes carry the different spin and move in the opposite direction.}
\label{fig:material}
\end{figure}

\textit{Material realization.}---We present the material realization of the valley contrasted quantum metric-based complete optical selection in the monolayer $\mathrm{V_2SeSO}$, which is the Janus derivative of $\mathrm{V_2Se_2O}$ and is a $d$-wave AM~\cite{xu2025chemical,ma2021multifunctional,bailing2024altermag}. Besides the AM family of $\mathrm{V_2Se_2O}$~\cite{xu2025chemical}, AM in twisted magnetic van der Waals materials~\cite{PhysRevLett.133.206702} also constitute a platform for realizing our proposed optical selection rules.

The crystal structure of monolayer $\mathrm{V_2SeSO}$ is shown in Fig.~\ref{fig:material}(a) and belongs to the space group $P4mm$ (No. 99). The two V atoms possess opposite magnetic moments and form the 2D $d$-wave AM order. Figure~\ref{fig:material}(b) shows the relevant mirror symmetries in monolayer $\mathrm{V_2SeSO}$. Due to the weak SOC in $\mathrm{V_2SeSO}$, the effects of SOC are neglected.

The monolayer $\mathrm{V_2SeSO}$ has two valleys at X and Y, and the two valleys are locked to opposite spins, as shown in Fig.~\ref{fig:material}(c). The two valleys are related by the mirror symmetries $M_x$ and $M_y$. Therefore, the degree of linear polarization $\eta(\boldsymbol{k})$ is opposite around the two valleys as shown in Fig.~\ref{fig:material}(d). Moreover, both the mirror-invariant lines of $M_{xy}$ and $M_{x,-y}$ pass through the two valleys. For the X valley, the conduction and valence bands have identical $M_{x,-y}$ eigenvalues but have different $M_{xy}$ eigenvalues. For the Y valley, the conduction and valence bands have identical $M_{xy}$ eigenvalues but have different $M_{x,-y}$ eigenvalues. From Eq.~(\ref{eq:mxy}) and Eq.~(\ref{eq:mx-y}), we have $g^{xy}_{vc}(\boldsymbol{k})=g^{xx}_{vc}(\boldsymbol{k})=g^{yy}_{vc}(\boldsymbol{k})$ for valley at X and $-g^{xy}_{vc}(\boldsymbol{k})=g^{xx}_{vc}(\boldsymbol{k})=g^{yy}_{vc}(\boldsymbol{k})$ for valley at Y. Therefore, $\eta(\mathrm{X})=-\eta(\mathrm{Y})=1$ according to Eq.~(\ref{eq:dop_twoband}). The $\hat{\boldsymbol{x}}+\hat{\boldsymbol{y}}$-polarized light is exclusively allowed at X valley and the $\hat{\boldsymbol{x}}-\hat{\boldsymbol{y}}$-polarized light is exclusively allowed at Y valley. As shown in Fig.~\ref{fig:material}(d), the overwhelming majority of optical transitions around the valleys preserve complete linear polarization selectivity.

The valley-contrasted optical selection rules based on the quantum metric in the monolayer $\mathrm{V_2SeSO}$ provide new pathways for spintronics and valleytronics applications. As shown in Fig.~\ref{fig:material}(e), an optical field with $\hat{\boldsymbol{x}}+\hat{\boldsymbol{y}}$ ($\hat{\boldsymbol{x}}-\hat{\boldsymbol{y}}$) polarization excites electrons at X (Y) valley. Due to the spin-valley locking between the X (Y) valley and the up (down) spin, the spin-up (spin-down) electrons are generated by linearly polarized light with $\hat{\boldsymbol{x}}+\hat{\boldsymbol{y}}$ ($\hat{\boldsymbol{x}}-\hat{\boldsymbol{y}}$) polarization. By applying an electric field to dissociate electron-hole pairs, fully spin- and valley-polarized currents can be generated.

\textit{Discussion.}---Conventional optical selection rules are rooted in the topology, i.e., the Berry curvature, and local atomic orbitals. However, the optical selection rules for linearly polarized light are believed to arise solely from local atomic orbitals.  We find that the geometry, i.e., the quantum metric, which has attracted significant attention in recent years~\cite{gao_field_2014, peotta_superfluidity_2015, torma2022superconductivity, Kaplan2024Unification, luhaizhou2024quantume, kang2025measurements}, can also serve as a fundamental origin for optical selection rules for linearly polarized light. Our quantum metric-based optical selection rules complete the geometric picture of optical selection rules. Both topology and geometry govern optical selection rules.

Just as the introduction of Berry curvature advanced our understanding of optical selection rules for circularly polarized light, our incorporation of the quantum metric also represents a fundamental advance for optical selection rules for linearly polarized light and enables new physics: 1. The direct experimental measurement of the quantum metric is a growing focus~\cite{kang_measurements_2025,kim_direct_2025}, and quantum metric-based optical selection rules indicate that one can directly estimate the quantum metric by the absorption difference for orthogonal linear polarizations. 2. Due to the significant enhancement of the stiffness of exciton condensates by the quantum metric~\cite{PhysRevB.105.L140506,PhysRevLett.132.236001}, we anticipate the remarkable exciton condensates under quantum metric-based optical selection. 3. The quantum metric-based optical selection rules make widely investigated AMs a good playground for optical selection rules, since a significant number of reported abundant AMs exhibit mirror or mirror-like symmetry~\cite{bai_altermagnetism_2024}. 4. The quantum metric-based optical selection rules unlock the potential of the $\Gamma$ valley without circular polarization selection for application in optical selection rules.

\begin{acknowledgments}
\textit{Acknowledgments.}---We thank Guibin Liu for helpful discussions. The work is supported by the NSF of China (Grant No. 12374055), the National Key R\&D Program of China (Grant No. 2020YFA0308800), and the Science Fund for Creative Research Groups of NSFC (Grant No. 12321004).
\end{acknowledgments}

\bibliography{reference}

\begin{thebibliography}{15}%
\makeatletter
\providecommand \@ifxundefined [1]{%
 \@ifx{#1\undefined}
}%
\providecommand \@ifnum [1]{%
 \ifnum #1\expandafter \@firstoftwo
 \else \expandafter \@secondoftwo
 \fi
}%
\providecommand \@ifx [1]{%
 \ifx #1\expandafter \@firstoftwo
 \else \expandafter \@secondoftwo
 \fi
}%
\providecommand \natexlab [1]{#1}%
\providecommand \enquote  [1]{``#1''}%
\providecommand \bibnamefont  [1]{#1}%
\providecommand \bibfnamefont [1]{#1}%
\providecommand \citenamefont [1]{#1}%
\providecommand \href@noop [0]{\@secondoftwo}%
\providecommand \href [0]{\begingroup \@sanitize@url \@href}%
\providecommand \@href[1]{\@@startlink{#1}\@@href}%
\providecommand \@@href[1]{\endgroup#1\@@endlink}%
\providecommand \@sanitize@url [0]{\catcode `\\12\catcode `\$12\catcode
  `\&12\catcode `\#12\catcode `\^12\catcode `\_12\catcode `\%12\relax}%
\providecommand \@@startlink[1]{}%
\providecommand \@@endlink[0]{}%
\providecommand \url  [0]{\begingroup\@sanitize@url \@url }%
\providecommand \@url [1]{\endgroup\@href {#1}{\urlprefix }}%
\providecommand \urlprefix  [0]{URL }%
\providecommand \Eprint [0]{\href }%
\providecommand \doibase [0]{https://doi.org/}%
\providecommand \selectlanguage [0]{\@gobble}%
\providecommand \bibinfo  [0]{\@secondoftwo}%
\providecommand \bibfield  [0]{\@secondoftwo}%
\providecommand \translation [1]{[#1]}%
\providecommand \BibitemOpen [0]{}%
\providecommand \bibitemStop [0]{}%
\providecommand \bibitemNoStop [0]{.\EOS\space}%
\providecommand \EOS [0]{\spacefactor3000\relax}%
\providecommand \BibitemShut  [1]{\csname bibitem#1\endcsname}%
\let\auto@bib@innerbib\@empty
\bibitem [{\citenamefont {Aversa}\ and\ \citenamefont
  {Sipe}(1995)}]{aversa_nonlinear_1995}%
  \BibitemOpen
  \bibfield  {author} {\bibinfo {author} {\bibfnamefont {C.}~\bibnamefont
  {Aversa}}\ and\ \bibinfo {author} {\bibfnamefont {J.~E.}\ \bibnamefont
  {Sipe}},\ }\bibfield  {title} {\bibinfo {title} {Nonlinear optical
  susceptibilities of semiconductors: {{Results}} with a length-gauge
  analysis},\ }\href {https://doi.org/10.1103/PhysRevB.52.14636} {\bibfield
  {journal} {\bibinfo  {journal} {Phys. Rev. B}\ }\textbf {\bibinfo {volume}
  {52}},\ \bibinfo {pages} {14636} (\bibinfo {year} {1995})}\BibitemShut
  {NoStop}%
\bibitem [{\citenamefont {Sakurai}(1994)}]{sakurai1994modern}%
  \BibitemOpen
  \bibfield  {author} {\bibinfo {author} {\bibfnamefont {J.}~\bibnamefont
  {Sakurai}},\ }\href@noop {} {\emph {\bibinfo {title} {Modern Quantum
  Physics}}}\ (\bibinfo  {publisher} {Addison-Wesley, Reading, MA},\ \bibinfo
  {year} {1994})\BibitemShut {NoStop}%
\bibitem [{\citenamefont {Yu}\ and\ \citenamefont
  {Cardona}(2010)}]{yu_fundamentals_2010a}%
  \BibitemOpen
  \bibfield  {author} {\bibinfo {author} {\bibfnamefont {P.~Y.}\ \bibnamefont
  {Yu}}\ and\ \bibinfo {author} {\bibfnamefont {M.}~\bibnamefont {Cardona}},\
  }\href@noop {} {\emph {\bibinfo {title} {Fundamentals of {{Semiconductors}}:
  {{Physics}} and {{Materials Properties}}}}}\ (\bibinfo  {publisher} {Springer
  Berlin Heidelberg},\ \bibinfo {year} {2010})\BibitemShut {NoStop}%
\bibitem [{\citenamefont {Souza}\ and\ \citenamefont
  {Vanderbilt}(2008)}]{souza_dichroic_2008}%
  \BibitemOpen
  \bibfield  {author} {\bibinfo {author} {\bibfnamefont {I.}~\bibnamefont
  {Souza}}\ and\ \bibinfo {author} {\bibfnamefont {D.}~\bibnamefont
  {Vanderbilt}},\ }\bibfield  {title} {\bibinfo {title} {Dichroic $f$-sum rule
  and the orbital magnetization of crystals},\ }\href
  {https://doi.org/10.1103/PhysRevB.77.054438} {\bibfield  {journal} {\bibinfo
  {journal} {Phys. Rev. B}\ }\textbf {\bibinfo {volume} {77}},\ \bibinfo
  {pages} {054438} (\bibinfo {year} {2008})}\BibitemShut {NoStop}%
\bibitem [{\citenamefont {Yao}\ \emph {et~al.}(2008)\citenamefont {Yao},
  \citenamefont {Xiao},\ and\ \citenamefont {Niu}}]{yao_valleydependent_2008}%
  \BibitemOpen
  \bibfield  {author} {\bibinfo {author} {\bibfnamefont {W.}~\bibnamefont
  {Yao}}, \bibinfo {author} {\bibfnamefont {D.}~\bibnamefont {Xiao}},\ and\
  \bibinfo {author} {\bibfnamefont {Q.}~\bibnamefont {Niu}},\ }\bibfield
  {title} {\bibinfo {title} {Valley-dependent optoelectronics from inversion
  symmetry breaking},\ }\href {https://doi.org/10.1103/PhysRevB.77.235406}
  {\bibfield  {journal} {\bibinfo  {journal} {Phys. Rev. B}\ }\textbf {\bibinfo
  {volume} {77}},\ \bibinfo {pages} {235406} (\bibinfo {year}
  {2008})}\BibitemShut {NoStop}%
\bibitem [{\citenamefont {Roy}(2014)}]{PhysRevB.90.165139}%
  \BibitemOpen
  \bibfield  {author} {\bibinfo {author} {\bibfnamefont {R.}~\bibnamefont
  {Roy}},\ }\bibfield  {title} {\bibinfo {title} {Band geometry of fractional
  topological insulators},\ }\href {https://doi.org/10.1103/PhysRevB.90.165139}
  {\bibfield  {journal} {\bibinfo  {journal} {Phys. Rev. B}\ }\textbf {\bibinfo
  {volume} {90}},\ \bibinfo {pages} {165139} (\bibinfo {year}
  {2014})}\BibitemShut {NoStop}%
\bibitem [{\citenamefont {Peotta}\ and\ \citenamefont
  {T{\"o}rm{\"a}}(2015)}]{peotta_superfluidity_2015}%
  \BibitemOpen
  \bibfield  {author} {\bibinfo {author} {\bibfnamefont {S.}~\bibnamefont
  {Peotta}}\ and\ \bibinfo {author} {\bibfnamefont {P.}~\bibnamefont
  {T{\"o}rm{\"a}}},\ }\bibfield  {title} {\bibinfo {title} {Superfluidity in
  topologically nontrivial flat bands},\ }\href
  {https://doi.org/10.1038/ncomms9944} {\bibfield  {journal} {\bibinfo
  {journal} {Nat. Commun.}\ }\textbf {\bibinfo {volume} {6}},\ \bibinfo {pages}
  {8944} (\bibinfo {year} {2015})}\BibitemShut {NoStop}%
\bibitem [{\citenamefont {Ozawa}\ and\ \citenamefont
  {Mera}(2021)}]{PhysRevB.104.045103}%
  \BibitemOpen
  \bibfield  {author} {\bibinfo {author} {\bibfnamefont {T.}~\bibnamefont
  {Ozawa}}\ and\ \bibinfo {author} {\bibfnamefont {B.}~\bibnamefont {Mera}},\
  }\bibfield  {title} {\bibinfo {title} {Relations between topology and the
  quantum metric for chern insulators},\ }\href
  {https://doi.org/10.1103/PhysRevB.104.045103} {\bibfield  {journal} {\bibinfo
   {journal} {Phys. Rev. B}\ }\textbf {\bibinfo {volume} {104}},\ \bibinfo
  {pages} {045103} (\bibinfo {year} {2021})}\BibitemShut {NoStop}%
\bibitem [{\citenamefont {Li}\ \emph {et~al.}(2025)\citenamefont {Li},
  \citenamefont {Liu},\ and\ \citenamefont {Liu}}]{li2025quantum}%
  \BibitemOpen
  \bibfield  {author} {\bibinfo {author} {\bibfnamefont {Y.}~\bibnamefont
  {Li}}, \bibinfo {author} {\bibfnamefont {Y.}~\bibnamefont {Liu}},\ and\
  \bibinfo {author} {\bibfnamefont {C.-C.}\ \bibnamefont {Liu}},\ }\bibfield
  {title} {\bibinfo {title} {Quantum metric induced magneto-optical effects in
  $\mathcal{PT}$-symmetric antiferromagnets},\ }\href
  {https://doi.org/10.48550/arXiv.2503.04312} {\bibfield  {journal} {\bibinfo
  {journal} {arXiv:2503.04312}\ } (\bibinfo {year} {2025})}\BibitemShut
  {NoStop}%
\bibitem [{\citenamefont {Kuch}\ \emph {et~al.}(2014)\citenamefont {Kuch},
  \citenamefont {Sch{\"a}fer}, \citenamefont {Fischer},\ and\ \citenamefont
  {{Franz Ulrich Hillebrecht}}}]{kuch_magnetic_2014}%
  \BibitemOpen
  \bibfield  {author} {\bibinfo {author} {\bibfnamefont {W.}~\bibnamefont
  {Kuch}}, \bibinfo {author} {\bibfnamefont {R.}~\bibnamefont {Sch{\"a}fer}},
  \bibinfo {author} {\bibfnamefont {P.}~\bibnamefont {Fischer}},\ and\ \bibinfo
  {author} {\bibnamefont {{Franz Ulrich Hillebrecht}}},\ }\href@noop {} {\emph
  {\bibinfo {title} {Magnetic {{Microscopy}} of {{Layered Structures}}}}}\
  (\bibinfo  {publisher} {Springer Berlin, Heidelberg},\ \bibinfo {year}
  {2014})\BibitemShut {NoStop}%
\bibitem [{\citenamefont {Perdew}\ \emph {et~al.}(1996)\citenamefont {Perdew},
  \citenamefont {Burke},\ and\ \citenamefont {Ernzerhof}}]{PBE}%
  \BibitemOpen
  \bibfield  {author} {\bibinfo {author} {\bibfnamefont {J.~P.}\ \bibnamefont
  {Perdew}}, \bibinfo {author} {\bibfnamefont {K.}~\bibnamefont {Burke}},\ and\
  \bibinfo {author} {\bibfnamefont {M.}~\bibnamefont {Ernzerhof}},\ }\bibfield
  {title} {\bibinfo {title} {Generalized gradient approximation made simple},\
  }\href {https://doi.org/10.1103/physrevlett.77.3865} {\bibfield  {journal}
  {\bibinfo  {journal} {Phys. Rev. Lett.}\ }\textbf {\bibinfo {volume} {77}},\
  \bibinfo {pages} {3865} (\bibinfo {year} {1996})}\BibitemShut {NoStop}%
\bibitem [{\citenamefont {Kresse}\ and\ \citenamefont
  {Furthm{\"u}ller}(1996)}]{vasp}%
  \BibitemOpen
  \bibfield  {author} {\bibinfo {author} {\bibfnamefont {G.}~\bibnamefont
  {Kresse}}\ and\ \bibinfo {author} {\bibfnamefont {J.}~\bibnamefont
  {Furthm{\"u}ller}},\ }\bibfield  {title} {\bibinfo {title} {Efficient
  iterative schemes for $ab$ initio total-energy calculations using a
  plane-wave basis set},\ }\href {https://doi.org/10.1103/physrevb.54.11169}
  {\bibfield  {journal} {\bibinfo  {journal} {Phys. Rev. B}\ }\textbf {\bibinfo
  {volume} {54}},\ \bibinfo {pages} {11169} (\bibinfo {year}
  {1996})}\BibitemShut {NoStop}%
\bibitem [{\citenamefont {Xu}\ \emph {et~al.}(2025)\citenamefont {Xu},
  \citenamefont {Gao},\ and\ \citenamefont {Liu}}]{xu2025chemical}%
  \BibitemOpen
  \bibfield  {author} {\bibinfo {author} {\bibfnamefont {R.}~\bibnamefont
  {Xu}}, \bibinfo {author} {\bibfnamefont {Y.}~\bibnamefont {Gao}},\ and\
  \bibinfo {author} {\bibfnamefont {J.}~\bibnamefont {Liu}},\ }\bibfield
  {title} {\bibinfo {title} {Chemical design of monolayer altermagnets},\
  }\href {https://doi.org/10.48550/arXiv.2505.15484} {\bibfield  {journal}
  {\bibinfo  {journal} {arXiv:2505.15484}\ } (\bibinfo {year}
  {2025})}\BibitemShut {NoStop}%
\bibitem [{\citenamefont {Mostofi}\ \emph {et~al.}(2008)\citenamefont
  {Mostofi}, \citenamefont {Yates}, \citenamefont {Lee}, \citenamefont {Souza},
  \citenamefont {Vanderbilt},\ and\ \citenamefont {Marzari}}]{wannier90}%
  \BibitemOpen
  \bibfield  {author} {\bibinfo {author} {\bibfnamefont {A.~A.}\ \bibnamefont
  {Mostofi}}, \bibinfo {author} {\bibfnamefont {J.~R.}\ \bibnamefont {Yates}},
  \bibinfo {author} {\bibfnamefont {Y.-S.}\ \bibnamefont {Lee}}, \bibinfo
  {author} {\bibfnamefont {I.}~\bibnamefont {Souza}}, \bibinfo {author}
  {\bibfnamefont {D.}~\bibnamefont {Vanderbilt}},\ and\ \bibinfo {author}
  {\bibfnamefont {N.}~\bibnamefont {Marzari}},\ }\bibfield  {title} {\bibinfo
  {title} {wannier90: A tool for obtaining maximally-localised wannier
  functions},\ }\href {https://doi.org/10.1016/j.cpc.2007.11.016} {\bibfield
  {journal} {\bibinfo  {journal} {Comput. Phys. Commun.}\ }\textbf {\bibinfo
  {volume} {178}},\ \bibinfo {pages} {685} (\bibinfo {year}
  {2008})}\BibitemShut {NoStop}%
\bibitem [{\citenamefont {Wang}\ \emph {et~al.}(2006)\citenamefont {Wang},
  \citenamefont {Yates}, \citenamefont {Souza},\ and\ \citenamefont
  {Vanderbilt}}]{wang_initio_2006}%
  \BibitemOpen
  \bibfield  {author} {\bibinfo {author} {\bibfnamefont {X.}~\bibnamefont
  {Wang}}, \bibinfo {author} {\bibfnamefont {J.~R.}\ \bibnamefont {Yates}},
  \bibinfo {author} {\bibfnamefont {I.}~\bibnamefont {Souza}},\ and\ \bibinfo
  {author} {\bibfnamefont {D.}~\bibnamefont {Vanderbilt}},\ }\bibfield  {title}
  {\bibinfo {title} {Ab initio calculation of the anomalous {{Hall}}
  conductivity by {{Wannier}} interpolation},\ }\href
  {https://doi.org/10.1103/PhysRevB.74.195118} {\bibfield  {journal} {\bibinfo
  {journal} {Phys. Rev. B}\ }\textbf {\bibinfo {volume} {74}},\ \bibinfo
  {pages} {195118} (\bibinfo {year} {2006})}\BibitemShut {NoStop}%
\end{thebibliography}%


\begin{thebibliography}{54}%
\makeatletter
\providecommand \@ifxundefined [1]{%
 \@ifx{#1\undefined}
}%
\providecommand \@ifnum [1]{%
 \ifnum #1\expandafter \@firstoftwo
 \else \expandafter \@secondoftwo
 \fi
}%
\providecommand \@ifx [1]{%
 \ifx #1\expandafter \@firstoftwo
 \else \expandafter \@secondoftwo
 \fi
}%
\providecommand \natexlab [1]{#1}%
\providecommand \enquote  [1]{``#1''}%
\providecommand \bibnamefont  [1]{#1}%
\providecommand \bibfnamefont [1]{#1}%
\providecommand \citenamefont [1]{#1}%
\providecommand \href@noop [0]{\@secondoftwo}%
\providecommand \href [0]{\begingroup \@sanitize@url \@href}%
\providecommand \@href[1]{\@@startlink{#1}\@@href}%
\providecommand \@@href[1]{\endgroup#1\@@endlink}%
\providecommand \@sanitize@url [0]{\catcode `\\12\catcode `\$12\catcode
  `\&12\catcode `\#12\catcode `\^12\catcode `\_12\catcode `\%12\relax}%
\providecommand \@@startlink[1]{}%
\providecommand \@@endlink[0]{}%
\providecommand \url  [0]{\begingroup\@sanitize@url \@url }%
\providecommand \@url [1]{\endgroup\@href {#1}{\urlprefix }}%
\providecommand \urlprefix  [0]{URL }%
\providecommand \Eprint [0]{\href }%
\providecommand \doibase [0]{https://doi.org/}%
\providecommand \selectlanguage [0]{\@gobble}%
\providecommand \bibinfo  [0]{\@secondoftwo}%
\providecommand \bibfield  [0]{\@secondoftwo}%
\providecommand \translation [1]{[#1]}%
\providecommand \BibitemOpen [0]{}%
\providecommand \bibitemStop [0]{}%
\providecommand \bibitemNoStop [0]{.\EOS\space}%
\providecommand \EOS [0]{\spacefactor3000\relax}%
\providecommand \BibitemShut  [1]{\csname bibitem#1\endcsname}%
\let\auto@bib@innerbib\@empty
\bibitem [{\citenamefont {Hasegawa}\ and\ \citenamefont
  {Howard}(1961)}]{hasegawa1961optical}%
  \BibitemOpen
  \bibfield  {author} {\bibinfo {author} {\bibfnamefont {H.}~\bibnamefont
  {Hasegawa}}\ and\ \bibinfo {author} {\bibfnamefont {R.}~\bibnamefont
  {Howard}},\ }\bibfield  {title} {\bibinfo {title} {Optical absorption
  spectrum of hydrogenic atoms in a strong magnetic field},\ }\href
  {https://doi.org/10.1016/0022-3697(61)90097-X} {\bibfield  {journal}
  {\bibinfo  {journal} {J. Phys. Chem. Solids}\ }\textbf {\bibinfo {volume}
  {21}},\ \bibinfo {pages} {179} (\bibinfo {year} {1961})}\BibitemShut
  {NoStop}%
\bibitem [{\citenamefont {Thonhauser}\ \emph {et~al.}(2005)\citenamefont
  {Thonhauser}, \citenamefont {Ceresoli}, \citenamefont {Vanderbilt},\ and\
  \citenamefont {Resta}}]{PhysRevLett.95.137205}%
  \BibitemOpen
  \bibfield  {author} {\bibinfo {author} {\bibfnamefont {T.}~\bibnamefont
  {Thonhauser}}, \bibinfo {author} {\bibfnamefont {D.}~\bibnamefont
  {Ceresoli}}, \bibinfo {author} {\bibfnamefont {D.}~\bibnamefont
  {Vanderbilt}},\ and\ \bibinfo {author} {\bibfnamefont {R.}~\bibnamefont
  {Resta}},\ }\bibfield  {title} {\bibinfo {title} {Orbital magnetization in
  periodic insulators},\ }\href {https://doi.org/10.1103/PhysRevLett.95.137205}
  {\bibfield  {journal} {\bibinfo  {journal} {Phys. Rev. Lett.}\ }\textbf
  {\bibinfo {volume} {95}},\ \bibinfo {pages} {137205} (\bibinfo {year}
  {2005})}\BibitemShut {NoStop}%
\bibitem [{\citenamefont {Xiao}\ \emph {et~al.}(2005)\citenamefont {Xiao},
  \citenamefont {Shi},\ and\ \citenamefont {Niu}}]{PhysRevLett.95.137204}%
  \BibitemOpen
  \bibfield  {author} {\bibinfo {author} {\bibfnamefont {D.}~\bibnamefont
  {Xiao}}, \bibinfo {author} {\bibfnamefont {J.}~\bibnamefont {Shi}},\ and\
  \bibinfo {author} {\bibfnamefont {Q.}~\bibnamefont {Niu}},\ }\bibfield
  {title} {\bibinfo {title} {Berry phase correction to electron density of
  states in solids},\ }\href {https://doi.org/10.1103/PhysRevLett.95.137204}
  {\bibfield  {journal} {\bibinfo  {journal} {Phys. Rev. Lett.}\ }\textbf
  {\bibinfo {volume} {95}},\ \bibinfo {pages} {137204} (\bibinfo {year}
  {2005})}\BibitemShut {NoStop}%
\bibitem [{\citenamefont {Ceresoli}\ \emph {et~al.}(2006)\citenamefont
  {Ceresoli}, \citenamefont {Thonhauser}, \citenamefont {Vanderbilt},\ and\
  \citenamefont {Resta}}]{PhysRevB.74.024408}%
  \BibitemOpen
  \bibfield  {author} {\bibinfo {author} {\bibfnamefont {D.}~\bibnamefont
  {Ceresoli}}, \bibinfo {author} {\bibfnamefont {T.}~\bibnamefont
  {Thonhauser}}, \bibinfo {author} {\bibfnamefont {D.}~\bibnamefont
  {Vanderbilt}},\ and\ \bibinfo {author} {\bibfnamefont {R.}~\bibnamefont
  {Resta}},\ }\bibfield  {title} {\bibinfo {title} {Orbital magnetization in
  crystalline solids: Multi-band insulators, chern insulators, and metals},\
  }\href {https://doi.org/10.1103/PhysRevB.74.024408} {\bibfield  {journal}
  {\bibinfo  {journal} {Phys. Rev. B}\ }\textbf {\bibinfo {volume} {74}},\
  \bibinfo {pages} {024408} (\bibinfo {year} {2006})}\BibitemShut {NoStop}%
\bibitem [{\citenamefont {Yao}\ \emph {et~al.}(2008)\citenamefont {Yao},
  \citenamefont {Xiao},\ and\ \citenamefont {Niu}}]{yao_valleydependent_2008}%
  \BibitemOpen
  \bibfield  {author} {\bibinfo {author} {\bibfnamefont {W.}~\bibnamefont
  {Yao}}, \bibinfo {author} {\bibfnamefont {D.}~\bibnamefont {Xiao}},\ and\
  \bibinfo {author} {\bibfnamefont {Q.}~\bibnamefont {Niu}},\ }\bibfield
  {title} {\bibinfo {title} {Valley-dependent optoelectronics from inversion
  symmetry breaking},\ }\href {https://doi.org/10.1103/PhysRevB.77.235406}
  {\bibfield  {journal} {\bibinfo  {journal} {Phys. Rev. B}\ }\textbf {\bibinfo
  {volume} {77}},\ \bibinfo {pages} {235406} (\bibinfo {year}
  {2008})}\BibitemShut {NoStop}%
\bibitem [{\citenamefont {Cao}\ \emph {et~al.}(2012)\citenamefont {Cao},
  \citenamefont {Wang}, \citenamefont {Han}, \citenamefont {Ye}, \citenamefont
  {Zhu}, \citenamefont {Shi}, \citenamefont {Niu}, \citenamefont {Tan},
  \citenamefont {Wang}, \citenamefont {Liu},\ and\ \citenamefont
  {Feng}}]{cao_valleyselective_2012}%
  \BibitemOpen
  \bibfield  {author} {\bibinfo {author} {\bibfnamefont {T.}~\bibnamefont
  {Cao}}, \bibinfo {author} {\bibfnamefont {G.}~\bibnamefont {Wang}}, \bibinfo
  {author} {\bibfnamefont {W.}~\bibnamefont {Han}}, \bibinfo {author}
  {\bibfnamefont {H.}~\bibnamefont {Ye}}, \bibinfo {author} {\bibfnamefont
  {C.}~\bibnamefont {Zhu}}, \bibinfo {author} {\bibfnamefont {J.}~\bibnamefont
  {Shi}}, \bibinfo {author} {\bibfnamefont {Q.}~\bibnamefont {Niu}}, \bibinfo
  {author} {\bibfnamefont {P.}~\bibnamefont {Tan}}, \bibinfo {author}
  {\bibfnamefont {E.}~\bibnamefont {Wang}}, \bibinfo {author} {\bibfnamefont
  {B.}~\bibnamefont {Liu}},\ and\ \bibinfo {author} {\bibfnamefont
  {J.}~\bibnamefont {Feng}},\ }\bibfield  {title} {\bibinfo {title}
  {Valley-selective circular dichroism of monolayer molybdenum disulphide},\
  }\href {https://doi.org/10.1038/ncomms1882} {\bibfield  {journal} {\bibinfo
  {journal} {Nat. Commun.}\ }\textbf {\bibinfo {volume} {3}},\ \bibinfo {pages}
  {887} (\bibinfo {year} {2012})}\BibitemShut {NoStop}%
\bibitem [{\citenamefont {Xiao}\ \emph {et~al.}(2012)\citenamefont {Xiao},
  \citenamefont {Liu}, \citenamefont {Feng}, \citenamefont {Xu},\ and\
  \citenamefont {Yao}}]{xiao_coupled_2012}%
  \BibitemOpen
  \bibfield  {author} {\bibinfo {author} {\bibfnamefont {D.}~\bibnamefont
  {Xiao}}, \bibinfo {author} {\bibfnamefont {G.-B.}\ \bibnamefont {Liu}},
  \bibinfo {author} {\bibfnamefont {W.}~\bibnamefont {Feng}}, \bibinfo {author}
  {\bibfnamefont {X.}~\bibnamefont {Xu}},\ and\ \bibinfo {author}
  {\bibfnamefont {W.}~\bibnamefont {Yao}},\ }\bibfield  {title} {\bibinfo
  {title} {Coupled {{Spin}} and {{Valley Physics}} in {{Monolayers}} of
  {{MoS}}$_2$ and {{Other Group-VI Dichalcogenides}}},\ }\href
  {https://doi.org/10.1103/PhysRevLett.108.196802} {\bibfield  {journal}
  {\bibinfo  {journal} {Phys. Rev. Lett.}\ }\textbf {\bibinfo {volume} {108}},\
  \bibinfo {pages} {196802} (\bibinfo {year} {2012})}\BibitemShut {NoStop}%
\bibitem [{\citenamefont {Provost}\ and\ \citenamefont
  {Vallee}(1980)}]{provost_riemannian_1980}%
  \BibitemOpen
  \bibfield  {author} {\bibinfo {author} {\bibfnamefont {J.~P.}\ \bibnamefont
  {Provost}}\ and\ \bibinfo {author} {\bibfnamefont {G.}~\bibnamefont
  {Vallee}},\ }\bibfield  {title} {\bibinfo {title} {Riemannian structure on
  manifolds of quantum states},\ }\href {https://doi.org/10.1007/BF02193559}
  {\bibfield  {journal} {\bibinfo  {journal} {Commun. Math. Phys.}\ }\textbf
  {\bibinfo {volume} {76}},\ \bibinfo {pages} {289} (\bibinfo {year}
  {1980})}\BibitemShut {NoStop}%
\bibitem [{\citenamefont {Shapere}\ and\ \citenamefont
  {Wilczek}(1989)}]{shapere1989geometric}%
  \BibitemOpen
  \bibfield  {author} {\bibinfo {author} {\bibfnamefont {A.}~\bibnamefont
  {Shapere}}\ and\ \bibinfo {author} {\bibfnamefont {F.}~\bibnamefont
  {Wilczek}},\ }\href@noop {} {\emph {\bibinfo {title} {Geometric Phases in
  Physics}}},\ Vol.~\bibinfo {volume} {5}\ (\bibinfo  {publisher} {World
  scientific},\ \bibinfo {year} {1989})\BibitemShut {NoStop}%
\bibitem [{\citenamefont {Ma}\ \emph {et~al.}(2010)\citenamefont {Ma},
  \citenamefont {Chen}, \citenamefont {Fan},\ and\ \citenamefont
  {Liu}}]{ma_abelian_2010}%
  \BibitemOpen
  \bibfield  {author} {\bibinfo {author} {\bibfnamefont {Y.-Q.}\ \bibnamefont
  {Ma}}, \bibinfo {author} {\bibfnamefont {S.}~\bibnamefont {Chen}}, \bibinfo
  {author} {\bibfnamefont {H.}~\bibnamefont {Fan}},\ and\ \bibinfo {author}
  {\bibfnamefont {W.-M.}\ \bibnamefont {Liu}},\ }\bibfield  {title} {\bibinfo
  {title} {Abelian and non-{{Abelian}} quantum geometric tensor},\ }\href
  {https://doi.org/10.1103/PhysRevB.81.245129} {\bibfield  {journal} {\bibinfo
  {journal} {Phys. Rev. B}\ }\textbf {\bibinfo {volume} {81}},\ \bibinfo
  {pages} {245129} (\bibinfo {year} {2010})}\BibitemShut {NoStop}%
\bibitem [{\citenamefont {Aversa}\ and\ \citenamefont
  {Sipe}(1995)}]{aversa_nonlinear_1995}%
  \BibitemOpen
  \bibfield  {author} {\bibinfo {author} {\bibfnamefont {C.}~\bibnamefont
  {Aversa}}\ and\ \bibinfo {author} {\bibfnamefont {J.~E.}\ \bibnamefont
  {Sipe}},\ }\bibfield  {title} {\bibinfo {title} {Nonlinear optical
  susceptibilities of semiconductors: {{Results}} with a length-gauge
  analysis},\ }\href {https://doi.org/10.1103/PhysRevB.52.14636} {\bibfield
  {journal} {\bibinfo  {journal} {Phys. Rev. B}\ }\textbf {\bibinfo {volume}
  {52}},\ \bibinfo {pages} {14636} (\bibinfo {year} {1995})}\BibitemShut
  {NoStop}%
\bibitem [{\citenamefont {De~Juan}\ \emph {et~al.}(2017)\citenamefont
  {De~Juan}, \citenamefont {Grushin}, \citenamefont {Morimoto},\ and\
  \citenamefont {Moore}}]{de2017quantized}%
  \BibitemOpen
  \bibfield  {author} {\bibinfo {author} {\bibfnamefont {F.}~\bibnamefont
  {De~Juan}}, \bibinfo {author} {\bibfnamefont {A.~G.}\ \bibnamefont
  {Grushin}}, \bibinfo {author} {\bibfnamefont {T.}~\bibnamefont {Morimoto}},\
  and\ \bibinfo {author} {\bibfnamefont {J.~E.}\ \bibnamefont {Moore}},\
  }\bibfield  {title} {\bibinfo {title} {Quantized circular photogalvanic
  effect in {Weyl} semimetals},\ }\href {https://doi.org/10.1038/ncomms15995}
  {\bibfield  {journal} {\bibinfo  {journal} {Nat. Commun.}\ }\textbf {\bibinfo
  {volume} {8}},\ \bibinfo {pages} {15995} (\bibinfo {year}
  {2017})}\BibitemShut {NoStop}%
\bibitem [{\citenamefont {Ahn}\ \emph {et~al.}(2022)\citenamefont {Ahn},
  \citenamefont {Guo}, \citenamefont {Nagaosa},\ and\ \citenamefont
  {Vishwanath}}]{ahn_riemannian_2022}%
  \BibitemOpen
  \bibfield  {author} {\bibinfo {author} {\bibfnamefont {J.}~\bibnamefont
  {Ahn}}, \bibinfo {author} {\bibfnamefont {G.-Y.}\ \bibnamefont {Guo}},
  \bibinfo {author} {\bibfnamefont {N.}~\bibnamefont {Nagaosa}},\ and\ \bibinfo
  {author} {\bibfnamefont {A.}~\bibnamefont {Vishwanath}},\ }\bibfield  {title}
  {\bibinfo {title} {Riemannian geometry of resonant optical responses},\
  }\href {https://doi.org/10.1038/s41567-021-01465-z} {\bibfield  {journal}
  {\bibinfo  {journal} {Nat. Phys.}\ }\textbf {\bibinfo {volume} {18}},\
  \bibinfo {pages} {290} (\bibinfo {year} {2022})}\BibitemShut {NoStop}%
\bibitem [{\citenamefont {Verma}\ and\ \citenamefont
  {Queiroz}(2024)}]{verma2024instantaneous}%
  \BibitemOpen
  \bibfield  {author} {\bibinfo {author} {\bibfnamefont {N.}~\bibnamefont
  {Verma}}\ and\ \bibinfo {author} {\bibfnamefont {R.}~\bibnamefont
  {Queiroz}},\ }\bibfield  {title} {\bibinfo {title} {Instantaneous response
  and quantum geometry of insulators},\ }\bibfield  {journal} {\bibinfo
  {journal} {arXiv:2403.07052}\ }\href
  {https://doi.org/10.48550/arXiv.2403.07052} {10.48550/arXiv.2403.07052}
  (\bibinfo {year} {2024})\BibitemShut {NoStop}%
\bibitem [{\citenamefont {Ezawa}(2024)}]{ezawa2024analytic}%
  \BibitemOpen
  \bibfield  {author} {\bibinfo {author} {\bibfnamefont {M.}~\bibnamefont
  {Ezawa}},\ }\bibfield  {title} {\bibinfo {title} {Analytic approach to
  quantum metric and optical conductivity in dirac models with parabolic mass
  in arbitrary dimensions},\ }\href
  {https://doi.org/10.1103/PhysRevB.110.195437} {\bibfield  {journal} {\bibinfo
   {journal} {Phys. Rev. B}\ }\textbf {\bibinfo {volume} {110}},\ \bibinfo
  {pages} {195437} (\bibinfo {year} {2024})}\BibitemShut {NoStop}%
\bibitem [{\citenamefont {Komissarov}\ \emph {et~al.}(2024)\citenamefont
  {Komissarov}, \citenamefont {Holder},\ and\ \citenamefont
  {Queiroz}}]{komissarov_quantum_2024}%
  \BibitemOpen
  \bibfield  {author} {\bibinfo {author} {\bibfnamefont {I.}~\bibnamefont
  {Komissarov}}, \bibinfo {author} {\bibfnamefont {T.}~\bibnamefont {Holder}},\
  and\ \bibinfo {author} {\bibfnamefont {R.}~\bibnamefont {Queiroz}},\
  }\bibfield  {title} {\bibinfo {title} {The quantum geometric origin of
  capacitance in insulators},\ }\href
  {https://doi.org/10.1038/s41467-024-48808-x} {\bibfield  {journal} {\bibinfo
  {journal} {Nat. Commun.}\ }\textbf {\bibinfo {volume} {15}},\ \bibinfo
  {pages} {4621} (\bibinfo {year} {2024})}\BibitemShut {NoStop}%
\bibitem [{\citenamefont {Wang}\ \emph {et~al.}(2022)\citenamefont {Wang},
  \citenamefont {Yu}, \citenamefont {R\"osner}, \citenamefont {Katsnelson},
  \citenamefont {Lin},\ and\ \citenamefont {Yuan}}]{PhysRevX.12.021055}%
  \BibitemOpen
  \bibfield  {author} {\bibinfo {author} {\bibfnamefont {Y.}~\bibnamefont
  {Wang}}, \bibinfo {author} {\bibfnamefont {G.}~\bibnamefont {Yu}}, \bibinfo
  {author} {\bibfnamefont {M.}~\bibnamefont {R\"osner}}, \bibinfo {author}
  {\bibfnamefont {M.~I.}\ \bibnamefont {Katsnelson}}, \bibinfo {author}
  {\bibfnamefont {H.-Q.}\ \bibnamefont {Lin}},\ and\ \bibinfo {author}
  {\bibfnamefont {S.}~\bibnamefont {Yuan}},\ }\bibfield  {title} {\bibinfo
  {title} {Polarization-dependent selection rules and optical spectrum atlas of
  twisted bilayer graphene quantum dots},\ }\href
  {https://doi.org/10.1103/PhysRevX.12.021055} {\bibfield  {journal} {\bibinfo
  {journal} {Phys. Rev. X}\ }\textbf {\bibinfo {volume} {12}},\ \bibinfo
  {pages} {021055} (\bibinfo {year} {2022})}\BibitemShut {NoStop}%
\bibitem [{\citenamefont {Rodin}\ \emph {et~al.}(2016)\citenamefont {Rodin},
  \citenamefont {Gomes}, \citenamefont {Carvalho},\ and\ \citenamefont
  {Castro~Neto}}]{PhysRevB.93.045431}%
  \BibitemOpen
  \bibfield  {author} {\bibinfo {author} {\bibfnamefont {A.~S.}\ \bibnamefont
  {Rodin}}, \bibinfo {author} {\bibfnamefont {L.~C.}\ \bibnamefont {Gomes}},
  \bibinfo {author} {\bibfnamefont {A.}~\bibnamefont {Carvalho}},\ and\
  \bibinfo {author} {\bibfnamefont {A.~H.}\ \bibnamefont {Castro~Neto}},\
  }\bibfield  {title} {\bibinfo {title} {Valley physics in tin (ii) sulfide},\
  }\href {https://doi.org/10.1103/PhysRevB.93.045431} {\bibfield  {journal}
  {\bibinfo  {journal} {Phys. Rev. B}\ }\textbf {\bibinfo {volume} {93}},\
  \bibinfo {pages} {045431} (\bibinfo {year} {2016})}\BibitemShut {NoStop}%
\bibitem [{\citenamefont {Hanakata}\ \emph {et~al.}(2016)\citenamefont
  {Hanakata}, \citenamefont {Carvalho}, \citenamefont {Campbell},\ and\
  \citenamefont {Park}}]{PhysRevB.94.035304}%
  \BibitemOpen
  \bibfield  {author} {\bibinfo {author} {\bibfnamefont {P.~Z.}\ \bibnamefont
  {Hanakata}}, \bibinfo {author} {\bibfnamefont {A.}~\bibnamefont {Carvalho}},
  \bibinfo {author} {\bibfnamefont {D.~K.}\ \bibnamefont {Campbell}},\ and\
  \bibinfo {author} {\bibfnamefont {H.~S.}\ \bibnamefont {Park}},\ }\bibfield
  {title} {\bibinfo {title} {Polarization and valley switching in monolayer
  group-iv monochalcogenides},\ }\href
  {https://doi.org/10.1103/PhysRevB.94.035304} {\bibfield  {journal} {\bibinfo
  {journal} {Phys. Rev. B}\ }\textbf {\bibinfo {volume} {94}},\ \bibinfo
  {pages} {035304} (\bibinfo {year} {2016})}\BibitemShut {NoStop}%
\bibitem [{\citenamefont {Chen}\ \emph {et~al.}(2018)\citenamefont {Chen},
  \citenamefont {Chen}, \citenamefont {Shao}, \citenamefont {Deng},
  \citenamefont {Guo}, \citenamefont {Ma},\ and\ \citenamefont
  {Xia}}]{chen2018valley}%
  \BibitemOpen
  \bibfield  {author} {\bibinfo {author} {\bibfnamefont {C.}~\bibnamefont
  {Chen}}, \bibinfo {author} {\bibfnamefont {X.}~\bibnamefont {Chen}}, \bibinfo
  {author} {\bibfnamefont {Y.}~\bibnamefont {Shao}}, \bibinfo {author}
  {\bibfnamefont {B.}~\bibnamefont {Deng}}, \bibinfo {author} {\bibfnamefont
  {Q.}~\bibnamefont {Guo}}, \bibinfo {author} {\bibfnamefont {C.}~\bibnamefont
  {Ma}},\ and\ \bibinfo {author} {\bibfnamefont {F.}~\bibnamefont {Xia}},\
  }\bibfield  {title} {\bibinfo {title} {Valley-selective linear dichroism in
  layered tin sulfide},\ }\href {https://doi.org/10.1021/acsphotonics.8b00850}
  {\bibfield  {journal} {\bibinfo  {journal} {ACS Photonics}\ }\textbf
  {\bibinfo {volume} {5}},\ \bibinfo {pages} {3814} (\bibinfo {year}
  {2018})}\BibitemShut {NoStop}%
\bibitem [{\citenamefont {Yu}\ \emph {et~al.}(2020)\citenamefont {Yu},
  \citenamefont {Guan}, \citenamefont {Sheng}, \citenamefont {Gao},\ and\
  \citenamefont {Yang}}]{PhysRevLett.124.037701}%
  \BibitemOpen
  \bibfield  {author} {\bibinfo {author} {\bibfnamefont {Z.-M.}\ \bibnamefont
  {Yu}}, \bibinfo {author} {\bibfnamefont {S.}~\bibnamefont {Guan}}, \bibinfo
  {author} {\bibfnamefont {X.-L.}\ \bibnamefont {Sheng}}, \bibinfo {author}
  {\bibfnamefont {W.}~\bibnamefont {Gao}},\ and\ \bibinfo {author}
  {\bibfnamefont {S.~A.}\ \bibnamefont {Yang}},\ }\bibfield  {title} {\bibinfo
  {title} {Valley-layer coupling: A new design principle for valleytronics},\
  }\href {https://doi.org/10.1103/PhysRevLett.124.037701} {\bibfield  {journal}
  {\bibinfo  {journal} {Phys. Rev. Lett.}\ }\textbf {\bibinfo {volume} {124}},\
  \bibinfo {pages} {037701} (\bibinfo {year} {2020})}\BibitemShut {NoStop}%
\bibitem [{\citenamefont {Vila}\ \emph {et~al.}(2025)\citenamefont {Vila},
  \citenamefont {Sunko},\ and\ \citenamefont {Moore}}]{bzzy-ngcs}%
  \BibitemOpen
  \bibfield  {author} {\bibinfo {author} {\bibfnamefont {M.}~\bibnamefont
  {Vila}}, \bibinfo {author} {\bibfnamefont {V.}~\bibnamefont {Sunko}},\ and\
  \bibinfo {author} {\bibfnamefont {J.~E.}\ \bibnamefont {Moore}},\ }\bibfield
  {title} {\bibinfo {title} {Orbital-spin locking and its optical signatures in
  altermagnets},\ }\href {https://doi.org/10.1103/bzzy-ngcs} {\bibfield
  {journal} {\bibinfo  {journal} {Phys. Rev. B}\ }\textbf {\bibinfo {volume}
  {112}},\ \bibinfo {pages} {L020401} (\bibinfo {year} {2025})}\BibitemShut
  {NoStop}%
\bibitem [{SM()}]{SM}%
  \BibitemOpen
  \href@noop {} {\bibinfo  {journal} {See Supplemental Material for more
  detailed information on (I) Detailed derivation of the optical selection
  rules, (II) Complete polarization selectivity from the perspective of
  eigenmodes of light at the valley, (III) More details about the degree of
  polarization in the Kane-Mele model, (IV) Calculation details, which includes
  Ref.
  \cite{VASP,wannier90,PBE,wang_initio_2006,aversa_nonlinear_1995,xu2025chemical,yu_fundamentals_2010a,sakurai1994modern,souza_dichroic_2008,yao_valleydependent_2008,peotta_superfluidity_2015,PhysRevB.90.165139,PhysRevB.104.045103,li2025quantum,kuch_magnetic_2014}}\
  }\BibitemShut {NoStop}%
\bibitem [{\citenamefont {Yu}\ and\ \citenamefont
  {Cardona}(2010)}]{yu_fundamentals_2010a}%
  \BibitemOpen
\bibfield  {journal} {  }\bibfield  {author} {\bibinfo {author} {\bibfnamefont
  {P.~Y.}\ \bibnamefont {Yu}}\ and\ \bibinfo {author} {\bibfnamefont
  {M.}~\bibnamefont {Cardona}},\ }\href@noop {} {\emph {\bibinfo {title}
  {Fundamentals of {{Semiconductors}}: {{Physics}} and {{Materials
  Properties}}}}}\ (\bibinfo  {publisher} {Springer Berlin Heidelberg},\
  \bibinfo {year} {2010})\BibitemShut {NoStop}%
\bibitem [{\citenamefont {Sakurai}\ and\ \citenamefont
  {Napolitano}(2020)}]{sakurai2020modern}%
  \BibitemOpen
  \bibfield  {author} {\bibinfo {author} {\bibfnamefont {J.~J.}\ \bibnamefont
  {Sakurai}}\ and\ \bibinfo {author} {\bibfnamefont {J.}~\bibnamefont
  {Napolitano}},\ }\href@noop {} {\emph {\bibinfo {title} {Modern quantum
  mechanics}}}\ (\bibinfo  {publisher} {Cambridge University Press},\ \bibinfo
  {year} {2020})\BibitemShut {NoStop}%
\bibitem [{\citenamefont {Souza}\ and\ \citenamefont
  {Vanderbilt}(2008)}]{souza_dichroic_2008}%
  \BibitemOpen
  \bibfield  {author} {\bibinfo {author} {\bibfnamefont {I.}~\bibnamefont
  {Souza}}\ and\ \bibinfo {author} {\bibfnamefont {D.}~\bibnamefont
  {Vanderbilt}},\ }\bibfield  {title} {\bibinfo {title} {Dichroic $f$-sum rule
  and the orbital magnetization of crystals},\ }\href
  {https://doi.org/10.1103/PhysRevB.77.054438} {\bibfield  {journal} {\bibinfo
  {journal} {Phys. Rev. B}\ }\textbf {\bibinfo {volume} {77}},\ \bibinfo
  {pages} {054438} (\bibinfo {year} {2008})}\BibitemShut {NoStop}%
\bibitem [{\citenamefont {Peotta}\ and\ \citenamefont
  {T{\"o}rm{\"a}}(2015)}]{peotta_superfluidity_2015}%
  \BibitemOpen
  \bibfield  {author} {\bibinfo {author} {\bibfnamefont {S.}~\bibnamefont
  {Peotta}}\ and\ \bibinfo {author} {\bibfnamefont {P.}~\bibnamefont
  {T{\"o}rm{\"a}}},\ }\bibfield  {title} {\bibinfo {title} {Superfluidity in
  topologically nontrivial flat bands},\ }\href
  {https://doi.org/10.1038/ncomms9944} {\bibfield  {journal} {\bibinfo
  {journal} {Nat. Commun.}\ }\textbf {\bibinfo {volume} {6}},\ \bibinfo {pages}
  {8944} (\bibinfo {year} {2015})}\BibitemShut {NoStop}%
\bibitem [{\citenamefont {Roy}(2014)}]{PhysRevB.90.165139}%
  \BibitemOpen
  \bibfield  {author} {\bibinfo {author} {\bibfnamefont {R.}~\bibnamefont
  {Roy}},\ }\bibfield  {title} {\bibinfo {title} {Band geometry of fractional
  topological insulators},\ }\href {https://doi.org/10.1103/PhysRevB.90.165139}
  {\bibfield  {journal} {\bibinfo  {journal} {Phys. Rev. B}\ }\textbf {\bibinfo
  {volume} {90}},\ \bibinfo {pages} {165139} (\bibinfo {year}
  {2014})}\BibitemShut {NoStop}%
\bibitem [{\citenamefont {Ozawa}\ and\ \citenamefont
  {Mera}(2021)}]{PhysRevB.104.045103}%
  \BibitemOpen
  \bibfield  {author} {\bibinfo {author} {\bibfnamefont {T.}~\bibnamefont
  {Ozawa}}\ and\ \bibinfo {author} {\bibfnamefont {B.}~\bibnamefont {Mera}},\
  }\bibfield  {title} {\bibinfo {title} {Relations between topology and the
  quantum metric for chern insulators},\ }\href
  {https://doi.org/10.1103/PhysRevB.104.045103} {\bibfield  {journal} {\bibinfo
   {journal} {Phys. Rev. B}\ }\textbf {\bibinfo {volume} {104}},\ \bibinfo
  {pages} {045103} (\bibinfo {year} {2021})}\BibitemShut {NoStop}%
\bibitem [{\citenamefont {{\v S}mejkal}\ \emph {et~al.}(2022)\citenamefont {{\v
  S}mejkal}, \citenamefont {Sinova},\ and\ \citenamefont
  {Jungwirth}}]{smejkal_conventional_2022}%
  \BibitemOpen
  \bibfield  {author} {\bibinfo {author} {\bibfnamefont {L.}~\bibnamefont {{\v
  S}mejkal}}, \bibinfo {author} {\bibfnamefont {J.}~\bibnamefont {Sinova}},\
  and\ \bibinfo {author} {\bibfnamefont {T.}~\bibnamefont {Jungwirth}},\
  }\bibfield  {title} {\bibinfo {title} {Beyond {{Conventional Ferromagnetism}}
  and {{Antiferromagnetism}}: {{A Phase}} with {{Nonrelativistic Spin}} and
  {{Crystal Rotation Symmetry}}},\ }\href
  {https://doi.org/10.1103/PhysRevX.12.031042} {\bibfield  {journal} {\bibinfo
  {journal} {Phys. Rev. X}\ }\textbf {\bibinfo {volume} {12}},\ \bibinfo
  {pages} {031042} (\bibinfo {year} {2022})}\BibitemShut {NoStop}%
\bibitem [{\citenamefont {Liu}\ \emph {et~al.}(2022)\citenamefont {Liu},
  \citenamefont {Li}, \citenamefont {Han}, \citenamefont {Wan},\ and\
  \citenamefont {Liu}}]{liu_spingroup_2022}%
  \BibitemOpen
  \bibfield  {author} {\bibinfo {author} {\bibfnamefont {P.}~\bibnamefont
  {Liu}}, \bibinfo {author} {\bibfnamefont {J.}~\bibnamefont {Li}}, \bibinfo
  {author} {\bibfnamefont {J.}~\bibnamefont {Han}}, \bibinfo {author}
  {\bibfnamefont {X.}~\bibnamefont {Wan}},\ and\ \bibinfo {author}
  {\bibfnamefont {Q.}~\bibnamefont {Liu}},\ }\bibfield  {title} {\bibinfo
  {title} {Spin-{{Group Symmetry}} in {{Magnetic Materials}} with {{Negligible
  Spin-Orbit Coupling}}},\ }\href {https://doi.org/10.1103/PhysRevX.12.021016}
  {\bibfield  {journal} {\bibinfo  {journal} {Phys. Rev. X}\ }\textbf {\bibinfo
  {volume} {12}},\ \bibinfo {pages} {021016} (\bibinfo {year}
  {2022})}\BibitemShut {NoStop}%
\bibitem [{\citenamefont {Kane}\ and\ \citenamefont
  {Mele}(2005{\natexlab{a}})}]{PhysRevLett.95.146802}%
  \BibitemOpen
  \bibfield  {author} {\bibinfo {author} {\bibfnamefont {C.~L.}\ \bibnamefont
  {Kane}}\ and\ \bibinfo {author} {\bibfnamefont {E.~J.}\ \bibnamefont
  {Mele}},\ }\bibfield  {title} {\bibinfo {title} {${Z}_{2}$ topological order
  and the quantum spin hall effect},\ }\href
  {https://doi.org/10.1103/PhysRevLett.95.146802} {\bibfield  {journal}
  {\bibinfo  {journal} {Phys. Rev. Lett.}\ }\textbf {\bibinfo {volume} {95}},\
  \bibinfo {pages} {146802} (\bibinfo {year} {2005}{\natexlab{a}})}\BibitemShut
  {NoStop}%
\bibitem [{\citenamefont {Kane}\ and\ \citenamefont
  {Mele}(2005{\natexlab{b}})}]{PhysRevLett.95.226801}%
  \BibitemOpen
  \bibfield  {author} {\bibinfo {author} {\bibfnamefont {C.~L.}\ \bibnamefont
  {Kane}}\ and\ \bibinfo {author} {\bibfnamefont {E.~J.}\ \bibnamefont
  {Mele}},\ }\bibfield  {title} {\bibinfo {title} {Quantum spin hall effect in
  graphene},\ }\href {https://doi.org/10.1103/PhysRevLett.95.226801} {\bibfield
   {journal} {\bibinfo  {journal} {Phys. Rev. Lett.}\ }\textbf {\bibinfo
  {volume} {95}},\ \bibinfo {pages} {226801} (\bibinfo {year}
  {2005}{\natexlab{b}})}\BibitemShut {NoStop}%
\bibitem [{\citenamefont {Xu}\ \emph {et~al.}(2025)\citenamefont {Xu},
  \citenamefont {Gao},\ and\ \citenamefont {Liu}}]{xu2025chemical}%
  \BibitemOpen
  \bibfield  {author} {\bibinfo {author} {\bibfnamefont {R.}~\bibnamefont
  {Xu}}, \bibinfo {author} {\bibfnamefont {Y.}~\bibnamefont {Gao}},\ and\
  \bibinfo {author} {\bibfnamefont {J.}~\bibnamefont {Liu}},\ }\bibfield
  {title} {\bibinfo {title} {Chemical design of monolayer altermagnets},\
  }\href {https://doi.org/10.48550/arXiv.2505.15484} {\bibfield  {journal}
  {\bibinfo  {journal} {arXiv:2505.15484}\ } (\bibinfo {year}
  {2025})}\BibitemShut {NoStop}%
\bibitem [{\citenamefont {Ma}\ \emph {et~al.}(2021)\citenamefont {Ma},
  \citenamefont {Hu}, \citenamefont {Li}, \citenamefont {Liu}, \citenamefont
  {Yao}, \citenamefont {Jia},\ and\ \citenamefont
  {Liu}}]{ma2021multifunctional}%
  \BibitemOpen
  \bibfield  {author} {\bibinfo {author} {\bibfnamefont {H.-Y.}\ \bibnamefont
  {Ma}}, \bibinfo {author} {\bibfnamefont {M.}~\bibnamefont {Hu}}, \bibinfo
  {author} {\bibfnamefont {N.}~\bibnamefont {Li}}, \bibinfo {author}
  {\bibfnamefont {J.}~\bibnamefont {Liu}}, \bibinfo {author} {\bibfnamefont
  {W.}~\bibnamefont {Yao}}, \bibinfo {author} {\bibfnamefont {J.-F.}\
  \bibnamefont {Jia}},\ and\ \bibinfo {author} {\bibfnamefont {J.}~\bibnamefont
  {Liu}},\ }\bibfield  {title} {\bibinfo {title} {Multifunctional
  antiferromagnetic materials with giant piezomagnetism and noncollinear spin
  current},\ }\href {https://doi.org/10.1038/s41467-021-23127-7} {\bibfield
  {journal} {\bibinfo  {journal} {Nat. Commun.}\ }\textbf {\bibinfo {volume}
  {12}},\ \bibinfo {pages} {2846} (\bibinfo {year} {2021})}\BibitemShut
  {NoStop}%
\bibitem [{\citenamefont {Bai}\ \emph {et~al.}(2024{\natexlab{a}})\citenamefont
  {Bai}, \citenamefont {Feng}, \citenamefont {Liu}, \citenamefont
  {{\v{S}}mejkal}, \citenamefont {Mokrousov},\ and\ \citenamefont
  {Yao}}]{bailing2024altermag}%
  \BibitemOpen
  \bibfield  {author} {\bibinfo {author} {\bibfnamefont {L.}~\bibnamefont
  {Bai}}, \bibinfo {author} {\bibfnamefont {W.}~\bibnamefont {Feng}}, \bibinfo
  {author} {\bibfnamefont {S.}~\bibnamefont {Liu}}, \bibinfo {author}
  {\bibfnamefont {L.}~\bibnamefont {{\v{S}}mejkal}}, \bibinfo {author}
  {\bibfnamefont {Y.}~\bibnamefont {Mokrousov}},\ and\ \bibinfo {author}
  {\bibfnamefont {Y.}~\bibnamefont {Yao}},\ }\bibfield  {title} {\bibinfo
  {title} {Altermagnetism: Exploring new frontiers in magnetism and
  spintronics},\ }\href {https://doi.org/10.1002/adfm.202409327} {\bibfield
  {journal} {\bibinfo  {journal} {Adv. Funct. Mater.}\ }\textbf {\bibinfo
  {volume} {34}},\ \bibinfo {pages} {2409327} (\bibinfo {year}
  {2024}{\natexlab{a}})}\BibitemShut {NoStop}%
\bibitem [{\citenamefont {Liu}\ \emph {et~al.}(2024{\natexlab{a}})\citenamefont
  {Liu}, \citenamefont {Yu},\ and\ \citenamefont
  {Liu}}]{PhysRevLett.133.206702}%
  \BibitemOpen
  \bibfield  {author} {\bibinfo {author} {\bibfnamefont {Y.}~\bibnamefont
  {Liu}}, \bibinfo {author} {\bibfnamefont {J.}~\bibnamefont {Yu}},\ and\
  \bibinfo {author} {\bibfnamefont {C.-C.}\ \bibnamefont {Liu}},\ }\bibfield
  {title} {\bibinfo {title} {Twisted magnetic van der waals bilayers: An ideal
  platform for altermagnetism},\ }\href
  {https://doi.org/10.1103/PhysRevLett.133.206702} {\bibfield  {journal}
  {\bibinfo  {journal} {Phys. Rev. Lett.}\ }\textbf {\bibinfo {volume} {133}},\
  \bibinfo {pages} {206702} (\bibinfo {year} {2024}{\natexlab{a}})}\BibitemShut
  {NoStop}%
\bibitem [{\citenamefont {Gao}\ \emph {et~al.}(2014)\citenamefont {Gao},
  \citenamefont {Yang},\ and\ \citenamefont {Niu}}]{gao_field_2014}%
  \BibitemOpen
  \bibfield  {author} {\bibinfo {author} {\bibfnamefont {Y.}~\bibnamefont
  {Gao}}, \bibinfo {author} {\bibfnamefont {S.~A.}\ \bibnamefont {Yang}},\ and\
  \bibinfo {author} {\bibfnamefont {Q.}~\bibnamefont {Niu}},\ }\bibfield
  {title} {\bibinfo {title} {Field induced positional shift of {Bloch}
  electrons and its dynamical implications},\ }\href
  {https://doi.org/10.1103/PhysRevLett.112.166601} {\bibfield  {journal}
  {\bibinfo  {journal} {Phys. Rev. Lett.}\ }\textbf {\bibinfo {volume} {112}},\
  \bibinfo {pages} {166601} (\bibinfo {year} {2014})}\BibitemShut {NoStop}%
\bibitem [{\citenamefont {T{\"o}rm{\"a}}\ \emph {et~al.}(2022)\citenamefont
  {T{\"o}rm{\"a}}, \citenamefont {Peotta},\ and\ \citenamefont
  {Bernevig}}]{torma2022superconductivity}%
  \BibitemOpen
  \bibfield  {author} {\bibinfo {author} {\bibfnamefont {P.}~\bibnamefont
  {T{\"o}rm{\"a}}}, \bibinfo {author} {\bibfnamefont {S.}~\bibnamefont
  {Peotta}},\ and\ \bibinfo {author} {\bibfnamefont {B.~A.}\ \bibnamefont
  {Bernevig}},\ }\bibfield  {title} {\bibinfo {title} {Superconductivity,
  superfluidity and quantum geometry in twisted multilayer systems},\ }\href
  {https://doi.org/10.1038/s42254-022-00466-y} {\bibfield  {journal} {\bibinfo
  {journal} {Nat. Rev. Phys.}\ }\textbf {\bibinfo {volume} {4}},\ \bibinfo
  {pages} {528} (\bibinfo {year} {2022})}\BibitemShut {NoStop}%
\bibitem [{\citenamefont {Kaplan}\ \emph {et~al.}(2024)\citenamefont {Kaplan},
  \citenamefont {Holder},\ and\ \citenamefont {Yan}}]{Kaplan2024Unification}%
  \BibitemOpen
  \bibfield  {author} {\bibinfo {author} {\bibfnamefont {D.}~\bibnamefont
  {Kaplan}}, \bibinfo {author} {\bibfnamefont {T.}~\bibnamefont {Holder}},\
  and\ \bibinfo {author} {\bibfnamefont {B.}~\bibnamefont {Yan}},\ }\bibfield
  {title} {\bibinfo {title} {Unification of nonlinear anomalous hall effect and
  nonreciprocal magnetoresistance in metals by the quantum geometry},\ }\href
  {https://doi.org/10.1103/PhysRevLett.132.026301} {\bibfield  {journal}
  {\bibinfo  {journal} {Phys. Rev. Lett.}\ }\textbf {\bibinfo {volume} {132}},\
  \bibinfo {pages} {026301} (\bibinfo {year} {2024})}\BibitemShut {NoStop}%
\bibitem [{\citenamefont {Liu}\ \emph {et~al.}(2024{\natexlab{b}})\citenamefont
  {Liu}, \citenamefont {Qiang}, \citenamefont {Lu},\ and\ \citenamefont
  {Xie}}]{luhaizhou2024quantume}%
  \BibitemOpen
  \bibfield  {author} {\bibinfo {author} {\bibfnamefont {T.}~\bibnamefont
  {Liu}}, \bibinfo {author} {\bibfnamefont {X.-B.}\ \bibnamefont {Qiang}},
  \bibinfo {author} {\bibfnamefont {H.-Z.}\ \bibnamefont {Lu}},\ and\ \bibinfo
  {author} {\bibfnamefont {X.~C.}\ \bibnamefont {Xie}},\ }\bibfield  {title}
  {\bibinfo {title} {Quantum geometry in condensed matter},\ }\href
  {https://doi.org/10.1093/nsr/nwae334} {\bibfield  {journal} {\bibinfo
  {journal} {Natl. Sci. Rev.}\ ,\ \bibinfo {pages} {nwae334}} (\bibinfo {year}
  {2024}{\natexlab{b}})}\BibitemShut {NoStop}%
\bibitem [{\citenamefont {Kang}\ \emph
  {et~al.}(2025{\natexlab{a}})\citenamefont {Kang}, \citenamefont {Kim},
  \citenamefont {Qian}, \citenamefont {Neves}, \citenamefont {Ye},
  \citenamefont {Jung}, \citenamefont {Puntel}, \citenamefont {Mazzola},
  \citenamefont {Fang}, \citenamefont {Jozwiak} \emph
  {et~al.}}]{kang2025measurements}%
  \BibitemOpen
  \bibfield  {author} {\bibinfo {author} {\bibfnamefont {M.}~\bibnamefont
  {Kang}}, \bibinfo {author} {\bibfnamefont {S.}~\bibnamefont {Kim}}, \bibinfo
  {author} {\bibfnamefont {Y.}~\bibnamefont {Qian}}, \bibinfo {author}
  {\bibfnamefont {P.~M.}\ \bibnamefont {Neves}}, \bibinfo {author}
  {\bibfnamefont {L.}~\bibnamefont {Ye}}, \bibinfo {author} {\bibfnamefont
  {J.}~\bibnamefont {Jung}}, \bibinfo {author} {\bibfnamefont {D.}~\bibnamefont
  {Puntel}}, \bibinfo {author} {\bibfnamefont {F.}~\bibnamefont {Mazzola}},
  \bibinfo {author} {\bibfnamefont {S.}~\bibnamefont {Fang}}, \bibinfo {author}
  {\bibfnamefont {C.}~\bibnamefont {Jozwiak}}, \emph {et~al.},\ }\bibfield
  {title} {\bibinfo {title} {Measurements of the quantum geometric tensor in
  solids},\ }\href {https://doi.org/10.1038/s41567-024-02678-8} {\bibfield
  {journal} {\bibinfo  {journal} {Nature Physics}\ }\textbf {\bibinfo {volume}
  {21}},\ \bibinfo {pages} {110} (\bibinfo {year}
  {2025}{\natexlab{a}})}\BibitemShut {NoStop}%
\bibitem [{\citenamefont {Kang}\ \emph
  {et~al.}(2025{\natexlab{b}})\citenamefont {Kang}, \citenamefont {Kim},
  \citenamefont {Qian}, \citenamefont {Neves}, \citenamefont {Ye},
  \citenamefont {Jung}, \citenamefont {Puntel}, \citenamefont {Mazzola},
  \citenamefont {Fang}, \citenamefont {Jozwiak}, \citenamefont {Bostwick},
  \citenamefont {Rotenberg}, \citenamefont {Fuji}, \citenamefont {Vobornik},
  \citenamefont {Park}, \citenamefont {Checkelsky}, \citenamefont {Yang},\ and\
  \citenamefont {Comin}}]{kang_measurements_2025}%
  \BibitemOpen
  \bibfield  {author} {\bibinfo {author} {\bibfnamefont {M.}~\bibnamefont
  {Kang}}, \bibinfo {author} {\bibfnamefont {S.}~\bibnamefont {Kim}}, \bibinfo
  {author} {\bibfnamefont {Y.}~\bibnamefont {Qian}}, \bibinfo {author}
  {\bibfnamefont {P.~M.}\ \bibnamefont {Neves}}, \bibinfo {author}
  {\bibfnamefont {L.}~\bibnamefont {Ye}}, \bibinfo {author} {\bibfnamefont
  {J.}~\bibnamefont {Jung}}, \bibinfo {author} {\bibfnamefont {D.}~\bibnamefont
  {Puntel}}, \bibinfo {author} {\bibfnamefont {F.}~\bibnamefont {Mazzola}},
  \bibinfo {author} {\bibfnamefont {S.}~\bibnamefont {Fang}}, \bibinfo {author}
  {\bibfnamefont {C.}~\bibnamefont {Jozwiak}}, \bibinfo {author} {\bibfnamefont
  {A.}~\bibnamefont {Bostwick}}, \bibinfo {author} {\bibfnamefont
  {E.}~\bibnamefont {Rotenberg}}, \bibinfo {author} {\bibfnamefont
  {J.}~\bibnamefont {Fuji}}, \bibinfo {author} {\bibfnamefont {I.}~\bibnamefont
  {Vobornik}}, \bibinfo {author} {\bibfnamefont {J.-H.}\ \bibnamefont {Park}},
  \bibinfo {author} {\bibfnamefont {J.~G.}\ \bibnamefont {Checkelsky}},
  \bibinfo {author} {\bibfnamefont {B.-J.}\ \bibnamefont {Yang}},\ and\
  \bibinfo {author} {\bibfnamefont {R.}~\bibnamefont {Comin}},\ }\bibfield
  {title} {\bibinfo {title} {Measurements of the quantum geometric tensor in
  solids},\ }\href {https://doi.org/10.1038/s41567-024-02678-8} {\bibfield
  {journal} {\bibinfo  {journal} {Nat. Phys.}\ }\textbf {\bibinfo {volume}
  {21}},\ \bibinfo {pages} {110} (\bibinfo {year}
  {2025}{\natexlab{b}})}\BibitemShut {NoStop}%
\bibitem [{\citenamefont {Kim}\ \emph {et~al.}(2025)\citenamefont {Kim},
  \citenamefont {Chung}, \citenamefont {Qian}, \citenamefont {Park},
  \citenamefont {Jozwiak}, \citenamefont {Rotenberg}, \citenamefont {Bostwick},
  \citenamefont {Kim},\ and\ \citenamefont {Yang}}]{kim_direct_2025}%
  \BibitemOpen
  \bibfield  {author} {\bibinfo {author} {\bibfnamefont {S.}~\bibnamefont
  {Kim}}, \bibinfo {author} {\bibfnamefont {Y.}~\bibnamefont {Chung}}, \bibinfo
  {author} {\bibfnamefont {Y.}~\bibnamefont {Qian}}, \bibinfo {author}
  {\bibfnamefont {S.}~\bibnamefont {Park}}, \bibinfo {author} {\bibfnamefont
  {C.}~\bibnamefont {Jozwiak}}, \bibinfo {author} {\bibfnamefont
  {E.}~\bibnamefont {Rotenberg}}, \bibinfo {author} {\bibfnamefont
  {A.}~\bibnamefont {Bostwick}}, \bibinfo {author} {\bibfnamefont {K.~S.}\
  \bibnamefont {Kim}},\ and\ \bibinfo {author} {\bibfnamefont {B.-J.}\
  \bibnamefont {Yang}},\ }\bibfield  {title} {\bibinfo {title} {Direct
  measurement of the quantum metric tensor in solids},\ }\href
  {https://doi.org/10.1126/science.ado6049} {\bibfield  {journal} {\bibinfo
  {journal} {Science}\ }\textbf {\bibinfo {volume} {388}},\ \bibinfo {pages}
  {1050} (\bibinfo {year} {2025})}\BibitemShut {NoStop}%
\bibitem [{\citenamefont {Hu}\ \emph {et~al.}(2022)\citenamefont {Hu},
  \citenamefont {Hyart}, \citenamefont {Pikulin},\ and\ \citenamefont
  {Rossi}}]{PhysRevB.105.L140506}%
  \BibitemOpen
  \bibfield  {author} {\bibinfo {author} {\bibfnamefont {X.}~\bibnamefont
  {Hu}}, \bibinfo {author} {\bibfnamefont {T.}~\bibnamefont {Hyart}}, \bibinfo
  {author} {\bibfnamefont {D.~I.}\ \bibnamefont {Pikulin}},\ and\ \bibinfo
  {author} {\bibfnamefont {E.}~\bibnamefont {Rossi}},\ }\bibfield  {title}
  {\bibinfo {title} {Quantum-metric-enabled exciton condensate in double
  twisted bilayer graphene},\ }\href
  {https://doi.org/10.1103/PhysRevB.105.L140506} {\bibfield  {journal}
  {\bibinfo  {journal} {Phys. Rev. B}\ }\textbf {\bibinfo {volume} {105}},\
  \bibinfo {pages} {L140506} (\bibinfo {year} {2022})}\BibitemShut {NoStop}%
\bibitem [{\citenamefont {Verma}\ \emph {et~al.}(2024)\citenamefont {Verma},
  \citenamefont {Guerci},\ and\ \citenamefont
  {Queiroz}}]{PhysRevLett.132.236001}%
  \BibitemOpen
  \bibfield  {author} {\bibinfo {author} {\bibfnamefont {N.}~\bibnamefont
  {Verma}}, \bibinfo {author} {\bibfnamefont {D.}~\bibnamefont {Guerci}},\ and\
  \bibinfo {author} {\bibfnamefont {R.}~\bibnamefont {Queiroz}},\ }\bibfield
  {title} {\bibinfo {title} {Geometric stiffness in interlayer exciton
  condensates},\ }\href {https://doi.org/10.1103/PhysRevLett.132.236001}
  {\bibfield  {journal} {\bibinfo  {journal} {Phys. Rev. Lett.}\ }\textbf
  {\bibinfo {volume} {132}},\ \bibinfo {pages} {236001} (\bibinfo {year}
  {2024})}\BibitemShut {NoStop}%
\bibitem [{\citenamefont {Bai}\ \emph {et~al.}(2024{\natexlab{b}})\citenamefont
  {Bai}, \citenamefont {Feng}, \citenamefont {Liu}, \citenamefont {{\v
  S}mejkal}, \citenamefont {Mokrousov},\ and\ \citenamefont
  {Yao}}]{bai_altermagnetism_2024}%
  \BibitemOpen
  \bibfield  {author} {\bibinfo {author} {\bibfnamefont {L.}~\bibnamefont
  {Bai}}, \bibinfo {author} {\bibfnamefont {W.}~\bibnamefont {Feng}}, \bibinfo
  {author} {\bibfnamefont {S.}~\bibnamefont {Liu}}, \bibinfo {author}
  {\bibfnamefont {L.}~\bibnamefont {{\v S}mejkal}}, \bibinfo {author}
  {\bibfnamefont {Y.}~\bibnamefont {Mokrousov}},\ and\ \bibinfo {author}
  {\bibfnamefont {Y.}~\bibnamefont {Yao}},\ }\bibfield  {title} {\bibinfo
  {title} {Altermagnetism: {{Exploring New Frontiers}} in {{Magnetism}} and
  {{Spintronics}}},\ }\href {https://doi.org/10.1002/adfm.202409327} {\bibfield
   {journal} {\bibinfo  {journal} {Adv Funct Materials}\ }\textbf {\bibinfo
  {volume} {34}},\ \bibinfo {pages} {2409327} (\bibinfo {year}
  {2024}{\natexlab{b}})}\BibitemShut {NoStop}%
\bibitem [{\citenamefont {Kresse}\ and\ \citenamefont
  {Furthm{\"u}ller}(1996)}]{VASP}%
  \BibitemOpen
  \bibfield  {author} {\bibinfo {author} {\bibfnamefont {G.}~\bibnamefont
  {Kresse}}\ and\ \bibinfo {author} {\bibfnamefont {J.}~\bibnamefont
  {Furthm{\"u}ller}},\ }\bibfield  {title} {\bibinfo {title} {Efficient
  iterative schemes for $ab$ initio total-energy calculations using a
  plane-wave basis set},\ }\href {https://doi.org/10.1103/physrevb.54.11169}
  {\bibfield  {journal} {\bibinfo  {journal} {Phys. Rev. B}\ }\textbf {\bibinfo
  {volume} {54}},\ \bibinfo {pages} {11169} (\bibinfo {year}
  {1996})}\BibitemShut {NoStop}%
\bibitem [{\citenamefont {Mostofi}\ \emph {et~al.}(2008)\citenamefont
  {Mostofi}, \citenamefont {Yates}, \citenamefont {Lee}, \citenamefont {Souza},
  \citenamefont {Vanderbilt},\ and\ \citenamefont {Marzari}}]{wannier90}%
  \BibitemOpen
  \bibfield  {author} {\bibinfo {author} {\bibfnamefont {A.~A.}\ \bibnamefont
  {Mostofi}}, \bibinfo {author} {\bibfnamefont {J.~R.}\ \bibnamefont {Yates}},
  \bibinfo {author} {\bibfnamefont {Y.-S.}\ \bibnamefont {Lee}}, \bibinfo
  {author} {\bibfnamefont {I.}~\bibnamefont {Souza}}, \bibinfo {author}
  {\bibfnamefont {D.}~\bibnamefont {Vanderbilt}},\ and\ \bibinfo {author}
  {\bibfnamefont {N.}~\bibnamefont {Marzari}},\ }\bibfield  {title} {\bibinfo
  {title} {wannier90: A tool for obtaining maximally-localised wannier
  functions},\ }\href {https://doi.org/10.1016/j.cpc.2007.11.016} {\bibfield
  {journal} {\bibinfo  {journal} {Comput. Phys. Commun.}\ }\textbf {\bibinfo
  {volume} {178}},\ \bibinfo {pages} {685} (\bibinfo {year}
  {2008})}\BibitemShut {NoStop}%
\bibitem [{\citenamefont {Perdew}\ \emph {et~al.}(1996)\citenamefont {Perdew},
  \citenamefont {Burke},\ and\ \citenamefont {Ernzerhof}}]{PBE}%
  \BibitemOpen
  \bibfield  {author} {\bibinfo {author} {\bibfnamefont {J.~P.}\ \bibnamefont
  {Perdew}}, \bibinfo {author} {\bibfnamefont {K.}~\bibnamefont {Burke}},\ and\
  \bibinfo {author} {\bibfnamefont {M.}~\bibnamefont {Ernzerhof}},\ }\bibfield
  {title} {\bibinfo {title} {Generalized gradient approximation made simple},\
  }\href {https://doi.org/10.1103/physrevlett.77.3865} {\bibfield  {journal}
  {\bibinfo  {journal} {Phys. Rev. Lett.}\ }\textbf {\bibinfo {volume} {77}},\
  \bibinfo {pages} {3865} (\bibinfo {year} {1996})}\BibitemShut {NoStop}%
\bibitem [{\citenamefont {Wang}\ \emph {et~al.}(2006)\citenamefont {Wang},
  \citenamefont {Yates}, \citenamefont {Souza},\ and\ \citenamefont
  {Vanderbilt}}]{wang_initio_2006}%
  \BibitemOpen
  \bibfield  {author} {\bibinfo {author} {\bibfnamefont {X.}~\bibnamefont
  {Wang}}, \bibinfo {author} {\bibfnamefont {J.~R.}\ \bibnamefont {Yates}},
  \bibinfo {author} {\bibfnamefont {I.}~\bibnamefont {Souza}},\ and\ \bibinfo
  {author} {\bibfnamefont {D.}~\bibnamefont {Vanderbilt}},\ }\bibfield  {title}
  {\bibinfo {title} {Ab initio calculation of the anomalous {{Hall}}
  conductivity by {{Wannier}} interpolation},\ }\href
  {https://doi.org/10.1103/PhysRevB.74.195118} {\bibfield  {journal} {\bibinfo
  {journal} {Phys. Rev. B}\ }\textbf {\bibinfo {volume} {74}},\ \bibinfo
  {pages} {195118} (\bibinfo {year} {2006})}\BibitemShut {NoStop}%
\bibitem [{\citenamefont {Sakurai}(1994)}]{sakurai1994modern}%
  \BibitemOpen
  \bibfield  {author} {\bibinfo {author} {\bibfnamefont {J.}~\bibnamefont
  {Sakurai}},\ }\href@noop {} {\emph {\bibinfo {title} {Modern Quantum
  Physics}}}\ (\bibinfo  {publisher} {Addison-Wesley, Reading, MA},\ \bibinfo
  {year} {1994})\BibitemShut {NoStop}%
\bibitem [{\citenamefont {Li}\ \emph {et~al.}(2025)\citenamefont {Li},
  \citenamefont {Liu},\ and\ \citenamefont {Liu}}]{li2025quantum}%
  \BibitemOpen
  \bibfield  {author} {\bibinfo {author} {\bibfnamefont {Y.}~\bibnamefont
  {Li}}, \bibinfo {author} {\bibfnamefont {Y.}~\bibnamefont {Liu}},\ and\
  \bibinfo {author} {\bibfnamefont {C.-C.}\ \bibnamefont {Liu}},\ }\bibfield
  {title} {\bibinfo {title} {Quantum metric induced magneto-optical effects in
  $\mathcal{PT}$-symmetric antiferromagnets},\ }\href
  {https://doi.org/10.48550/arXiv.2503.04312} {\bibfield  {journal} {\bibinfo
  {journal} {arXiv:2503.04312}\ } (\bibinfo {year} {2025})}\BibitemShut
  {NoStop}%
\bibitem [{\citenamefont {Kuch}\ \emph {et~al.}(2014)\citenamefont {Kuch},
  \citenamefont {Sch{\"a}fer}, \citenamefont {Fischer},\ and\ \citenamefont
  {{Franz Ulrich Hillebrecht}}}]{kuch_magnetic_2014}%
  \BibitemOpen
  \bibfield  {author} {\bibinfo {author} {\bibfnamefont {W.}~\bibnamefont
  {Kuch}}, \bibinfo {author} {\bibfnamefont {R.}~\bibnamefont {Sch{\"a}fer}},
  \bibinfo {author} {\bibfnamefont {P.}~\bibnamefont {Fischer}},\ and\ \bibinfo
  {author} {\bibnamefont {{Franz Ulrich Hillebrecht}}},\ }\href@noop {} {\emph
  {\bibinfo {title} {Magnetic {{Microscopy}} of {{Layered Structures}}}}}\
  (\bibinfo  {publisher} {Springer Berlin, Heidelberg},\ \bibinfo {year}
  {2014})\BibitemShut {NoStop}%
\end{thebibliography}%
\end{document}


\title{Supplementary Materials for ``Quantum metric-based optical selection rules''}

\author{Yongpan Li}
\affiliation{Centre for Quantum Physics, Key Laboratory of Advanced Optoelectronic Quantum Architecture and Measurement (MOE), School of Physics, Beijing Institute of Technology, Beijing, 100081, China}
 
\author{Cheng-Cheng Liu}
\email{ccliu@bit.edu.cn}
\affiliation{Centre for Quantum Physics, Key Laboratory of Advanced Optoelectronic Quantum Architecture and Measurement (MOE), School of Physics, Beijing Institute of Technology, Beijing, 100081, China}

\maketitle

\tableofcontents

\section{Detailed derivation of the optical selection rules}

In this section, we present a detailed derivation of the optical selection rules based on Berry curvature and quantum metric.

\subsection{Oscillator strength}
The optical conductivity tensor $\boldsymbol{\sigma}$ can be given by~\cite{aversa_nonlinear_1995}
\begin{equation}\label{eq:oct}
\sigma_{ab}(\omega)=-\frac{ie^2}{\hbar}\sum_{n\neq m}\int\frac{d^3k}{(2\pi)^3}\frac{(f_{n\boldsymbol{k}}^{\mathrm{FD}}-f_{m\boldsymbol{k}}^{\mathrm{FD}})}{\omega_{nm\boldsymbol{k}}}\frac{v_{nm\boldsymbol{k}}^av_{mn\boldsymbol{k}}^b}{\omega_{nm\boldsymbol{k}}+\omega+i\eta}.
\end{equation}
$f_{n\boldsymbol{k}}^{\mathrm{FD}}$ is the Fermi-Dirac distribution function, $\hbar\omega_{nm\boldsymbol{k}}$ is the energy difference between the $n$th band and $m$th band at the same $\boldsymbol{k}$ point, and $\eta>0$ is a smearing parameter. $\boldsymbol{v}_{nm\boldsymbol{k}}$ is the matrix element of the velocity operator at $\boldsymbol{k}$ point.

The dissipative (or Hermitian) part is given by
\begin{equation}
\sigma_{ab}^{\mathrm{H}}(\omega)=-\frac{\pi e^2}{\hbar}\sum_{n\neq m}\int\frac{d^3k}{(2\pi)^3}(f_{n\boldsymbol{k}}^{\mathrm{FD}}-f_{m\boldsymbol{k}}^{\mathrm{FD}})\frac{v_{nm\boldsymbol{k}}^av_{mn\boldsymbol{k}}^b}{\omega_{nm\boldsymbol{k}}}\delta(\omega_{nm\boldsymbol{k}}+\omega).
\end{equation}
For $a=b$, the above equation can be rewritten as
\begin{equation}
\begin{aligned}
\sigma_{aa}^{\mathrm{H}}(\omega)=&-\frac{\pi e^2}{\hbar}\sum_{n\neq m}\int\frac{d^3k}{(2\pi)^3}(f_{n\boldsymbol{k}}^{\mathrm{FD}}-f_{m\boldsymbol{k}}^{\mathrm{FD}})\frac{v_{nm\boldsymbol{k}}^av_{mn\boldsymbol{k}}^a}{\omega_{nm\boldsymbol{k}}}\delta(\omega_{nm\boldsymbol{k}}+\omega)\\
=&-\frac{\pi e^2}{2m_e}\sum_{n\neq m}\int\frac{d^3k}{(2\pi)^3}(f_{n\boldsymbol{k}}^{\mathrm{FD}}-f_{m\boldsymbol{k}}^{\mathrm{FD}})f_{nm\boldsymbol{k}}^{(\hat{\boldsymbol{a}})}\delta(\omega_{nm\boldsymbol{k}}+\omega).
\end{aligned}
\end{equation}
Here $f_{nm\boldsymbol{k}}^{(\hat{\boldsymbol{a}})}$ is the oscillator strength for the transition between states $n$ and $m$ and is given by~\cite{sakurai1994modern,yu_fundamentals_2010a}
\begin{equation}\label{eq:os_st}
f_{nm\boldsymbol{k}}^{(\hat{\boldsymbol{\epsilon}})}=\dfrac{2m_e|\langle n|\hat{\boldsymbol{\epsilon}}\cdot\boldsymbol{v}_{\boldsymbol{k}}|m\rangle|^2}{\hbar\omega_{nm\boldsymbol{k}}},
\end{equation}
where $m_e$ is the electron mass and $|\langle n|\hat{\boldsymbol{\epsilon}}\cdot\boldsymbol{v}_{\boldsymbol{k}}|m\rangle|^2=\langle n|\hat{\boldsymbol{\epsilon}}\cdot\boldsymbol{v}|m\rangle\langle m|\hat{\boldsymbol{\epsilon}}\cdot\boldsymbol{v}|n\rangle$. This expression is valid for the linear polarization $\hat{\boldsymbol{\epsilon}}\in\{\hat{\boldsymbol{x}},\hat{\boldsymbol{y}}\}$ of light.

Then the oscillator strength is generalized to circularly polarized light, i.e., $\hat{\boldsymbol{\epsilon}}\in\{\hat{\boldsymbol{x}}+i\hat{\boldsymbol{y}},\hat{\boldsymbol{x}}-i\hat{\boldsymbol{y}}\}$, and reads~\cite{souza_dichroic_2008,yao_valleydependent_2008}
\begin{subequations}\label{eq:os_st_cpl}
\begin{align}
f_{nm\boldsymbol{k}}^{(\hat{\boldsymbol{x}}+i\hat{\boldsymbol{y}})}&=\dfrac{2m_e}{\hbar\omega_{nm\boldsymbol{k}}}\langle n\boldsymbol{k}|(v^x+iv^y)|m\boldsymbol{k}\rangle\langle m\boldsymbol{k}|(v^x-iv^y)|n\boldsymbol{k}\rangle,\\
f_{nm\boldsymbol{k}}^{(\hat{\boldsymbol{x}}-i\hat{\boldsymbol{y}})}&=\dfrac{2m_e}{\hbar\omega_{nm\boldsymbol{k}}}\langle n\boldsymbol{k}|(v^x-iv^y)|m\boldsymbol{k}\rangle\langle m\boldsymbol{k}|(v^x+iv^y)|n\boldsymbol{k}\rangle.
\end{align}
\end{subequations}
In this work, the oscillator strengths for linearly polarized light with $\hat{\boldsymbol{x}}+\hat{\boldsymbol{y}}$ polarization and $\hat{\boldsymbol{x}}-\hat{\boldsymbol{y}}$ polarization are used and are given by
\begin{subequations}\label{eq:os_st_lpl}
\begin{align}
f_{nm\boldsymbol{k}}^{(\hat{\boldsymbol{x}}+\hat{\boldsymbol{y}})}&=\dfrac{2m_e}{\hbar\omega_{nm\boldsymbol{k}}}\langle n\boldsymbol{k}|(v^x+v^y)|m\boldsymbol{k}\rangle\langle m\boldsymbol{k}|(v^x+v^y)|n\boldsymbol{k}\rangle,\\
f_{nm\boldsymbol{k}}^{(\hat{\boldsymbol{x}}-\hat{\boldsymbol{y}})}&=\dfrac{2m_e}{\hbar\omega_{nm\boldsymbol{k}}}\langle n\boldsymbol{k}|(v^x-v^y)|m\boldsymbol{k}\rangle\langle m\boldsymbol{k}|(v^x-v^y)|n\boldsymbol{k}\rangle.
\end{align}
\end{subequations}

\subsection{Quantum geometry}

The quantum metric tensor $\boldsymbol{g}$ and Berry curvature tensor $\boldsymbol{\Omega}$ are the real and imaginary parts of the quantum geometry tensor $\boldsymbol{Q}=\boldsymbol{g}-i\boldsymbol{\Omega}/2$, respectively, and are given by
\begin{subequations}
\begin{align}
\boldsymbol{Q}_{nm}=&\begin{pmatrix}Q_{nm}^{xx}&Q_{nm}^{xy}\\Q_{nm}^{yx}&Q_{nm}^{yy}\end{pmatrix}=\begin{pmatrix}g_{nm}^{xx}&g_{nm}^{xy}-\dfrac{i}{2}\Omega_{nm}^{xy}\\g_{nm}^{xy}+\dfrac{i}{2}\Omega_{nm}^{xy}&g_{nm}^{yy}\end{pmatrix},\\
g^{ab}_{nm}(\boldsymbol{k})&\equiv \dfrac{1}{2}\dfrac{v^a_{nm,\boldsymbol{k}}v^b_{mn,\boldsymbol{k}}+v^b_{nm,\boldsymbol{k}}v^a_{mn,\boldsymbol{k}}}{\omega_{nm\boldsymbol{k}}^2},\\
\Omega^{ab}_{nm}(\boldsymbol{k})&\equiv i\dfrac{v^a_{nm,\boldsymbol{k}}v^b_{mn,\boldsymbol{k}}-v^b_{nm,\boldsymbol{k}}v^a_{mn,\boldsymbol{k}}}{\omega_{nm\boldsymbol{k}}^2},
\end{align}
\end{subequations}
where we have used $g_{nm}^{xy}=g_{nm}^{yx}$ and $\Omega_{nm}^{xy}=-\Omega_{nm}^{yx}$.

Here we derive the relations between the oscillator strengths and the quantum geometry tensor. The oscillator strengths for circularly polarized light, i.e., Eq. (\ref{eq:os_st_cpl}), can be rewritten as
\begin{equation}
\begin{aligned}
f_{nm\boldsymbol{k}}^{(\hat{\boldsymbol{x}}+i\hat{\boldsymbol{y}})}=&\dfrac{2m_e}{\hbar\omega_{nm\boldsymbol{k}}}(v^x_{nm\boldsymbol{k}}+iv^y_{nm\boldsymbol{k}})(v^x_{mn\boldsymbol{k}}-iv^y_{mn\boldsymbol{k}})\\
=&\dfrac{2m_e}{\hbar\omega_{nm\boldsymbol{k}}}\left(v^x_{nm\boldsymbol{k}}v^x_{mn\boldsymbol{k}}+v^y_{nm\boldsymbol{k}}v^y_{mn\boldsymbol{k}}-iv^x_{nm\boldsymbol{k}}v^y_{mn\boldsymbol{k}}+iv^y_{nm\boldsymbol{k}}v^x_{nm\boldsymbol{k}}\right).
\end{aligned}
\end{equation}
\begin{equation}
\begin{aligned}
f_{nm\boldsymbol{k}}^{(\hat{\boldsymbol{x}}-i\hat{\boldsymbol{y}})}=&\dfrac{2m_e}{\hbar\omega_{nm\boldsymbol{k}}}(v^x_{nm\boldsymbol{k}}-iv^y_{nm\boldsymbol{k}})(v^x_{mn\boldsymbol{k}}+iv^y_{mn\boldsymbol{k}})\\
=&\dfrac{2m_e}{\hbar\omega_{nm\boldsymbol{k}}}\left(v^x_{nm\boldsymbol{k}}v^x_{mn\boldsymbol{k}}+v^y_{nm\boldsymbol{k}}v^y_{mn\boldsymbol{k}}+iv^x_{nm\boldsymbol{k}}v^y_{mn\boldsymbol{k}}-iv^y_{nm\boldsymbol{k}}v^x_{nm\boldsymbol{k}}\right).
\end{aligned}
\end{equation}
Therefore,
\begin{equation}
\begin{aligned}
f_{nm\boldsymbol{k}}^{(\hat{\boldsymbol{x}}+i\hat{\boldsymbol{y}})}-f_{nm\boldsymbol{k}}^{(\hat{\boldsymbol{x}}-i\hat{\boldsymbol{y}})}=&-i\dfrac{4m_e}{\hbar\omega_{nm\boldsymbol{k}}}\left(v^x_{nm\boldsymbol{k}}v^y_{mn\boldsymbol{k}}-v^y_{nm\boldsymbol{k}}v^x_{nm\boldsymbol{k}}\right)\\
=&-\dfrac{4m_e\omega_{nm\boldsymbol{k}}}{\hbar}\Omega^{xy}_{nm}(\boldsymbol{k}).
\end{aligned}
\end{equation}
The off-diagonal term of the Berry curvature tensor is related to the difference in oscillator strengths for left- and right-handed circularly polarized light.

The oscillator strengths for linearly polarized light with $\hat{\boldsymbol{x}}+\hat{\boldsymbol{y}}$ polarization and $\hat{\boldsymbol{x}}-\hat{\boldsymbol{y}}$ polarization, i.e., Eq. (\ref{eq:os_st_lpl}), can be rewritten as
\begin{equation}
\begin{aligned}
f_{nm\boldsymbol{k}}^{(\hat{\boldsymbol{x}}+\hat{\boldsymbol{y}})}=&\dfrac{2m_e}{\hbar\omega_{nm\boldsymbol{k}}}(v^x_{nm\boldsymbol{k}}+v^y_{nm\boldsymbol{k}})(v^x_{mn\boldsymbol{k}}+v^y_{mn\boldsymbol{k}})\\
=&\dfrac{2m_e}{\hbar\omega_{nm\boldsymbol{k}}}\left(v^x_{nm\boldsymbol{k}}v^x_{mn\boldsymbol{k}}+v^y_{nm\boldsymbol{k}}v^y_{mn\boldsymbol{k}}+v^x_{nm\boldsymbol{k}}v^y_{mn\boldsymbol{k}}+v^y_{nm\boldsymbol{k}}v^x_{nm\boldsymbol{k}}\right).
\end{aligned}
\end{equation}
\begin{equation}
\begin{aligned}
f_{nm\boldsymbol{k}}^{(\hat{\boldsymbol{x}}-\hat{\boldsymbol{y}})}=&\dfrac{2m_e}{\hbar\omega_{nm\boldsymbol{k}}}(v^x_{nm\boldsymbol{k}}-v^y_{nm\boldsymbol{k}})(v^x_{mn\boldsymbol{k}}-v^y_{mn\boldsymbol{k}})\\
=&\dfrac{2m_e}{\hbar\omega_{nm\boldsymbol{k}}}\left(v^x_{nm\boldsymbol{k}}v^x_{mn\boldsymbol{k}}+v^y_{nm\boldsymbol{k}}v^y_{mn\boldsymbol{k}}-v^x_{nm\boldsymbol{k}}v^y_{mn\boldsymbol{k}}-v^y_{nm\boldsymbol{k}}v^x_{nm\boldsymbol{k}}\right).
\end{aligned}
\end{equation}
Therefore,
\begin{equation}
\begin{aligned}
f_{nm\boldsymbol{k}}^{(\hat{\boldsymbol{x}}+\hat{\boldsymbol{y}})}-f_{nm\boldsymbol{k}}^{(\hat{\boldsymbol{x}}-\hat{\boldsymbol{y}})}=&\dfrac{4m_e}{\hbar\omega_{nm\boldsymbol{k}}}\left(v^x_{nm\boldsymbol{k}}v^y_{mn\boldsymbol{k}}+v^y_{nm\boldsymbol{k}}v^x_{nm\boldsymbol{k}}\right)\\
=&\dfrac{8m_e\omega_{nm\boldsymbol{k}}}{\hbar}g^{xy}_{nm}(\boldsymbol{k})
\end{aligned}
\end{equation}
The off-diagonal term of the quantum metric tensor is related to the difference in oscillator strengths for linearly polarized light with $\hat{\boldsymbol{x}}+\hat{\boldsymbol{y}}$ polarization and $\hat{\boldsymbol{x}}-\hat{\boldsymbol{y}}$ polarization.

The sum of oscillator strengths, regardless of whether for circularly [Eq. (\ref{eq:os_st_cpl})] or linearly polarized light [Eq. \ref{eq:os_st_lpl}], is connected to the diagonal terms of the quantum metric tensor.
\begin{equation}
\begin{aligned}
f_{nm\boldsymbol{k}}^{(\hat{\boldsymbol{x}}+\hat{\boldsymbol{y}})}+f_{nm\boldsymbol{k}}^{(\hat{\boldsymbol{x}}-\hat{\boldsymbol{y}})}=&f_{nm\boldsymbol{k}}^{(\hat{\boldsymbol{x}}+i\hat{\boldsymbol{y}})}+f_{nm\boldsymbol{k}}^{(\hat{\boldsymbol{x}}-i\hat{\boldsymbol{y}})}\\
=&\dfrac{4m_e}{\hbar\omega_{nm\boldsymbol{k}}}\left(v^x_{nm\boldsymbol{k}}v^x_{mn\boldsymbol{k}}+v^y_{nm\boldsymbol{k}}v^y_{nm\boldsymbol{k}}\right)\\
=&\dfrac{4m_e\omega_{nm\boldsymbol{k}}}{\hbar}(g^{xx}_{nm}(\boldsymbol{k})+g^{yy}_{nm}(\boldsymbol{k})).
\end{aligned}
\end{equation}

\subsection{Degree of polarization}
The degree of polarization for transition between states $n$ and $m$ is given by~\cite{souza_dichroic_2008,yao_valleydependent_2008}
\begin{equation}
\eta_{nm}(\boldsymbol{k})=\dfrac{f_{nm\boldsymbol{k}}^{(\hat{\boldsymbol{\epsilon}}_1)}-f_{nm\boldsymbol{k}}^{(\hat{\boldsymbol{\epsilon}}_2)}}{f_{nm\boldsymbol{k}}^{(\hat{\boldsymbol{\epsilon}}_1)}+f_{nm\boldsymbol{k}}^{(\hat{\boldsymbol{\epsilon}}_2)}},
\end{equation}
where $\hat{\boldsymbol{\epsilon}}_1=\hat{\boldsymbol{x}}+i\hat{\boldsymbol{y}},\hat{\boldsymbol{\epsilon}}_2=\hat{\boldsymbol{x}}-i\hat{\boldsymbol{y}}$ for the circularly polarized light, and $\hat{\boldsymbol{\epsilon}}_1=\hat{\boldsymbol{x}}+\hat{\boldsymbol{y}},\hat{\boldsymbol{\epsilon}}_2=\hat{\boldsymbol{x}}-\hat{\boldsymbol{y}}$ for the linearly polarized light.

For circularly polarized light [Eq. (\ref{eq:os_st_cpl})], the degree of circular polarization reads
\begin{equation}\label{eq:deg_circ_pola}
\eta_{nm}^{\mathrm{CPL}}(\boldsymbol{k})=-\frac{\Omega^{xy}_{nm}(\boldsymbol{k})}{g^{xx}_{nm}(\boldsymbol{k})+g^{yy}_{nm}(\boldsymbol{k})}.
\end{equation}
For linearly polarized light [Eq. (\ref{eq:os_st_lpl})], the degree of linear polarization reads
\begin{equation}\label{eq:deg_linear_pola}
\eta_{nm}^{\mathrm{LPL}}(\boldsymbol{k})=\frac{2g^{xy}_{nm}(\boldsymbol{k})}{g^{xx}_{nm}(\boldsymbol{k})+g^{yy}_{nm}(\boldsymbol{k})}.
\end{equation}

\subsection{Complete polarization selectivity}

Here, we demonstrate that the complete polarization selectivity, either circular or linear, can be achieved in two-band valleys.

The quantum geometry tensor $\boldsymbol{Q}$ is positive-semidefinite and satisfies~\cite{PhysRevB.90.165139,peotta_superfluidity_2015,PhysRevB.104.045103}
\begin{equation}
Q^{xx}(\boldsymbol{k})Q^{yy}(\boldsymbol{k})-Q^{xy}(\boldsymbol{k})Q^{yx}(\boldsymbol{k})\ge0,
\end{equation}
which can be rewritten in terms of the quantum metric tensor and Berry curvature tensor as
\begin{equation}\label{eq:smdf_qm}
g^{xx}(\boldsymbol{k})g^{yy}(\boldsymbol{k})-g^{xy}(\boldsymbol{k})^2-\frac{\Omega^{xy}(\boldsymbol{k})^2}{4}\ge0.
\end{equation}
Here,
\begin{subequations}
\begin{align}
Q^{ab}(\boldsymbol{k})=&\sum_{n\neq m}Q_{nm}^{ab}(\boldsymbol{k}),\\
g^{ab}(\boldsymbol{k})=&\sum_{n\neq m}g_{nm}^{ab}(\boldsymbol{k}),\\
\Omega^{ab}(\boldsymbol{k})=&\sum_{n\neq m}\Omega_{nm}^{ab}(\boldsymbol{k}).
\end{align}
\end{subequations}

\subsubsection{$n$-fold rotational symmetry ($n\ge3$)}\label{sec:rot_sym}

A symmetry operator $\mathcal{O}$ acts on a tensor $T$ as
\begin{equation}\label{eq:sym_cstr}
\mathcal{O}T^{ab}=\sum_{\alpha\beta}R(\mathcal{O})_{\alpha a}R(\mathcal{O})_{\beta b}T^{\alpha\beta},
\end{equation}
where $a,b,\alpha,\beta$ are Cartesian coordinates, $R(\mathcal{O})$ is the matrix representation of $\mathcal{O}$.

Assuming the $\boldsymbol{k}'$ point possesses three-fold rotational symmetry $C_{3z}$, whose matrix representation is given by
\begin{equation}
R(C_{3z})=\begin{pmatrix}\dfrac{1}{2}&\dfrac{\sqrt{3}}{2}\\-\dfrac{\sqrt{3}}{2}&\dfrac{1}{2}\end{pmatrix}.
\end{equation}
At $\boldsymbol{k}'$ point, by solving the system of equations (\ref{eq:sym_cstr}), we find that the quantum metric tensor satisfies $g^{xx}(\boldsymbol{k}')=g^{yy}(\boldsymbol{k}')$ and $g^{xy}(\boldsymbol{k}')=-g^{yx}(\boldsymbol{k}')$ and the Berry curvature tensor satisfies $\Omega^{xy}(\boldsymbol{k}')=-\Omega^{yx}(\boldsymbol{k}')$. Since $g^{xy}(\boldsymbol{k})$ is symmetric under swapping of $x$ and $y$, i.e., $g^{xy}(\boldsymbol{k})=g^{yx}(\boldsymbol{k})$, we have $g^{xy}(\boldsymbol{k}')=g^{yx}(\boldsymbol{k}')=0$ at $\boldsymbol{k}'$ point. Therefore, at $\boldsymbol{k}'$ point, Eq. (\ref{eq:smdf_qm}) becomes
\begin{equation}\label{eq:ineq_circ}
g^{xx}(\boldsymbol{k}')^2-\frac{\Omega^{xy}(\boldsymbol{k}')^2}{4}\ge0.
\end{equation}
The above inequality is saturated only for two-band systems~\cite{PhysRevB.90.165139,peotta_superfluidity_2015,PhysRevB.104.045103} and when $\Omega^{xy}(\boldsymbol{k}')/2=\pm g^{xx}(\boldsymbol{k}')$. When Eq. (\ref{eq:ineq_circ}) is saturated, the degree of circular polarization $\eta^{\mathrm{CPL}}(\boldsymbol{k}')$ [Eq. (\ref{eq:deg_circ_pola})] becomes
\begin{equation}
\eta^{\mathrm{CPL}}(\boldsymbol{k}')=-\frac{\Omega^{xy}(\boldsymbol{k}')}{g^{xx}(\boldsymbol{k}')+g^{yy}(\boldsymbol{k}')}=-\frac{\pm 2g^{xx}(\boldsymbol{k}')}{2g^{xx}(\boldsymbol{k}')}=\pm1,
\end{equation}
i.e., the optical transition at $\boldsymbol{k}'$ exclusively couples to left or right circularly polarized light.

Through similar steps, one also can find that $\eta^{\mathrm{CPL}}(\boldsymbol{k}')=\pm1$ for $\boldsymbol{k}'$ at the four-fold and six-fold rotation axes.

\subsubsection{Mirror symmetry}\label{sec:mirror_sym}
Without loss of generality, here we take the mirror symmetry $M_{xy}:(x,y)\to(-y,-x)$ as an example. Assuming the $\boldsymbol{k}'$ point locates at the mirror invariant plane of the mirror symmetry $M_{xy}$, whose matrix representation is given by
\begin{equation}
R(M_{xy})=\begin{pmatrix}0&-1\\-1&0\end{pmatrix}.
\end{equation}
At $\boldsymbol{k}'$ point, by solving the system of equations (\ref{eq:sym_cstr}), we find that the quantum metric tensor satisfies $g^{xx}(\boldsymbol{k}')=g^{yy}(\boldsymbol{k}')$ and $g^{xy}(\boldsymbol{k}')=g^{yx}(\boldsymbol{k}')$ and the Berry curvature tensor satisfies $\Omega^{xy}(\boldsymbol{k}')=\Omega^{yx}(\boldsymbol{k}')$. Since $\Omega^{xy}(\boldsymbol{k})$ is antisymmetric under swapping of $x$ and $y$, i.e., $\Omega^{xy}(\boldsymbol{k})=-\Omega^{yx}(\boldsymbol{k})$, we have $\Omega^{xy}(\boldsymbol{k}')=\Omega^{yx}(\boldsymbol{k}')=0$ at $\boldsymbol{k}'$ point. Therefore, at $\boldsymbol{k}'$ point, Eq. (\ref{eq:smdf_qm}) becomes
\begin{equation}\label{eq:ineq_linear}
g^{xx}(\boldsymbol{k}')^2-g^{xy}(\boldsymbol{k}')^2\ge0.
\end{equation}
The above inequality is saturated only for two-band systems~\cite{PhysRevB.90.165139,peotta_superfluidity_2015,PhysRevB.104.045103} and when $g^{xy}(\boldsymbol{k}')=\pm g^{xx}(\boldsymbol{k}')$. When Eq. (\ref{eq:ineq_linear}) is saturated, the degree of circular polarization $\eta^{\mathrm{LPL}}(\boldsymbol{k}')$ [Eq. (\ref{eq:deg_linear_pola})] becomes
\begin{equation}
\eta^{\mathrm{LPL}}(\boldsymbol{k}')=\frac{2g^{xy}(\boldsymbol{k}')}{g^{xx}(\boldsymbol{k}')+g^{yy}(\boldsymbol{k}')}=\frac{\pm2g^{xx}(\boldsymbol{k}')}{2g^{xx}(\boldsymbol{k}')}=\pm1,
\end{equation}
i.e., the optical transition at $\boldsymbol{k}'$ exclusively couples to linearly polarized light with $\hat{\boldsymbol{x}}+\hat{\boldsymbol{y}}$ or $\hat{\boldsymbol{x}}-\hat{\boldsymbol{y}}$.

\section{Complete polarization selectivity from the perspective of eigenmodes of light at the valley}

The optical conductivity tensor (\ref{eq:oct}) can be rewritten as
\begin{equation}\label{eq:oct_qg}
\begin{aligned}
\sigma_{ba}(\omega)=&\sigma_{ba}^g(\omega)+\sigma_{ba}^\Omega(\omega),\\
\sigma_{ba}^g(\omega)=&-\frac{i e^2}{\hbar}\sum_{n\neq m}\int_{\boldsymbol{k}}\frac{(f_{n\boldsymbol{k}}^{\mathrm{FD}}-f_{m\boldsymbol{k}}^{\mathrm{FD}})\omega_{nm\boldsymbol{k}}}{\omega_{nm\boldsymbol{k}}+\omega}g_{nm}^{ba}(\boldsymbol{k}),\\
\sigma_{ba}^\Omega(\omega)=&-\frac{e^2}{2\hbar}\sum_{n\neq m}\int_{\boldsymbol{k}}\frac{(f_{n\boldsymbol{k}}^{\mathrm{FD}}-f_{m\boldsymbol{k}}^{\mathrm{FD}})\omega_{nm\boldsymbol{k}}}{\omega_{nm\boldsymbol{k}}+\omega}\Omega_{nm}^{ba}(\boldsymbol{k}).
\end{aligned}
\end{equation}
Consider the scenario of an optical transition involving two nondegenerate energy levels $E_c$ and $E_v$ at the $\boldsymbol{k'}$ valley. The optical conductivity tensor can be simplified to
\begin{equation}
\begin{aligned}
\sigma_{ba}(\omega)=&\sigma_{ba}^g(\omega)+\sigma_{ba}^\Omega(\omega),\\
\sigma_{ba}^g(\omega)=&-\frac{i e^2}{\hbar}\frac{(f_{c\boldsymbol{k'}}^{\mathrm{FD}}-f_{v\boldsymbol{k'}}^{\mathrm{FD}})\omega_{cv\boldsymbol{k'}}}{\omega_{cv\boldsymbol{k'}}+\omega}g_{cv}^{ba}(\boldsymbol{k}'),\\
\sigma_{ba}^\Omega(\omega)=&-\frac{e^2}{2\hbar}\frac{(f_{c\boldsymbol{k'}}^{\mathrm{FD}}-f_{v\boldsymbol{k'}}^{\mathrm{FD}})\omega_{cv\boldsymbol{k'}}}{\omega_{cv\boldsymbol{k'}}+\omega}\Omega_{cv}^{ba}(\boldsymbol{k}').
\end{aligned}
\end{equation}

In terms of the optical conductivity tensor (\ref{eq:oct_qg}), the wave equations are given by~\cite{li2025quantum,kuch_magnetic_2014}
\begin{equation}\label{eq:wave_eq}
\begin{pmatrix}-k_z^2+k_0^2+ik_0^2\sigma_{xx}^g/\omega\epsilon_0&ik_0^2(\sigma_{xy}^g+\sigma_{xy}^\Omega)/\omega\epsilon_0\\ik_0^2(\sigma_{xy}^g-\sigma_{xy}^\Omega)/\omega\epsilon_0&-k_z^2+k_0^2+ik_0^2\sigma_{yy}^g/\omega\epsilon_0\end{pmatrix}\begin{pmatrix}E_x\\E_y\end{pmatrix}=0.
\end{equation}

\subsection{$n$-fold rotational symmetry ($n\ge3$)}

Under $n$-fold rotational symmetry ($n\ge3$), we have $g^{xx}_{cv}(\boldsymbol{k'})=g^{yy}_{cv}(\boldsymbol{k'})=\pm\Omega^{xy}_{cv}(\boldsymbol{k}')/2$ and $g^{xy}(\boldsymbol{k})=0$ (Sec. \ref{sec:rot_sym}). Therefore, the optical conductivity tensor (\ref{eq:oct_qg}) satisfied $\sigma_{xx}^g=\sigma_{yy}^g=\pm i\sigma_{xy}^\Omega=\mp i\sigma_{yx}^\Omega$ and $\sigma_{xy}^g=\sigma_{yx}^g=0$.

For $\sigma_{xx}^g=\sigma_{yy}^g=i\sigma_{xy}^\Omega=-i\sigma_{yx}^\Omega$ and $\sigma_{xy}^g=\sigma_{yx}^g=0$, the wave equations (\ref{eq:wave_eq}) becomes
\begin{equation}
\begin{pmatrix}-k_z^2+k_0^2-k_0^2\sigma_{xy}^\Omega/\omega\epsilon_0&ik_0^2\sigma_{xy}^\Omega/\omega\epsilon_0\\-ik_0^2\sigma_{xy}^\Omega/\omega\epsilon_0&-k_z^2+k_0^2-k_0^2\sigma_{xy}^\Omega/\omega\epsilon_0\end{pmatrix}\begin{pmatrix}E_x\\E_y\end{pmatrix}=0.
\end{equation}
Solving the wave equations, the eigenmodes of light at the valley $\boldsymbol{E}_\pm$ and the corresponding refractive indices $n_\pm \equiv k_z/k_0$ are
\begin{align}
\boldsymbol{E}_{\pm}=&\frac{\boldsymbol{x}\pm i\boldsymbol{y}}{\sqrt{2}}E_0e^{i(k_zz-\omega t)},
\end{align}
\begin{equation}
n_\pm=\sqrt{1-\frac{2i\sigma_{xy}^\Omega\pm 2i\sigma_{xy}^\Omega}{2i\omega\epsilon_0}}.
\end{equation}
The right circularly polarized light $E_+$ has a complex refractive index $n$, whereas the left circularly polarized light $E_-$ has a real refractive index $n=1$. Since the imaginary part of the refractive index represents optical absorption by the valley, the valley exclusively absorbs the right circularly polarized light, i.e., the valley exhibits complete circular polarization selectivity for the right circularly polarized light. For $\sigma_{xx}^g=\sigma_{yy}^g=-i\sigma_{xy}^\Omega=i\sigma_{yx}^\Omega$ and $\sigma_{xy}^g=\sigma_{yx}^g=0$, a similar analysis shows that the valley possesses complete circular polarization selectivity for the left circularly polarized light.

\subsection{Mirror symmetry}

Under mirror symmetry, we have $g^{xx}(\boldsymbol{k}')=g^{yy}(\boldsymbol{k}')=\pm g^{xy}(\boldsymbol{k}')=\pm g^{yx}(\boldsymbol{k}')$ and $\Omega^{xy}(\boldsymbol{k}')=-\Omega^{yx}(\boldsymbol{k}')=0$ (Sec.~\ref{sec:mirror_sym}). Therefore, the optical conductivity tensor (\ref{eq:oct_qg}) satisfied $\sigma_{xx}^g=\sigma_{yy}^g=\pm \sigma_{xy}^g=\pm \sigma_{yx}^g$ and $\sigma_{xy}^\Omega=\sigma_{yx}^\Omega=0$.

For $\sigma_{xx}^g=\sigma_{yy}^g=\sigma_{xy}^g=\sigma_{yx}^g$ and $\sigma_{xy}^\Omega=\sigma_{yx}^\Omega=0$, the wave equations (\ref{eq:wave_eq}) becomes
\begin{equation}\label{eq:wave_eq}
\begin{pmatrix}-k_z^2+k_0^2+ik_0^2\sigma_{xy}^g/\omega\epsilon_0&ik_0^2\sigma_{xy}^g/\omega\epsilon_0\\ik_0^2\sigma_{xy}^g/\omega\epsilon_0&-k_z^2+k_0^2+ik_0^2\sigma_{xx}^g/\omega\epsilon_0\end{pmatrix}\begin{pmatrix}E_x\\E_y\end{pmatrix}=0.
\end{equation}
Solving the wave equations, the eigenmodes of light at the valley $\boldsymbol{E}_{x\pm y}$ and the corresponding refractive indices $n_{x\pm y}$ are
\begin{align}
\boldsymbol{E}_{{x\pm y}}=&\frac{\boldsymbol{x}\pm \boldsymbol{y}}{\sqrt{2}}E_0e^{i(k_zz-\omega t)},
\end{align}
\begin{equation}
n_{x\pm y}=\sqrt{1-\frac{2\sigma_{xy}^g\pm 2\sigma_{xy}^g}{2i\omega\epsilon_0}}.
\end{equation}
The linearly polarized light polarized along the $\boldsymbol{x}+\boldsymbol{y}$ direction has a complex refractive index $n$, whereas the linearly polarized light polarized along the $\boldsymbol{x}-\boldsymbol{y}$ direction has a real refractive index $n=1$. The valley exclusively absorbs $\boldsymbol{x}+\boldsymbol{y}$-polarized light, i.e., the valley exhibits complete linear polarization selectivity for the $\boldsymbol{x}+\boldsymbol{y}$-polarized light. When $\sigma_{xx}^g=\sigma_{yy}^g=-\sigma_{xy}^g=-\sigma_{yx}^g$ and $\sigma_{xy}^\Omega=-\sigma_{yx}^\Omega=0$, a similar analysis shows that the valley possesses complete linear polarization selectivity for the $\boldsymbol{x}-\boldsymbol{y}$-polarized light.

\section{More details about the degree of polarization in the Kane-Mele model}

We rotate the lattice clockwise by 15$^{\circ}$, i.e., the angle between the KM high-symmetry line and the $x$-axis becomes 45$^{\circ}$ and the two mirror symmetries at M become $M_{xy}$ and $M_{x,-y}$. The $k\cdot p$ Hamiltonian at M becomes
\begin{eqnarray}
H_{\mathrm{KM,M}}&=&-t\left(\frac{1}{2}\sigma_xs_0+\frac{\sqrt{3}}{2}\sigma_ys_0\right)-t(k_x+k_y)\left(-\frac{1}{\sqrt{2}}\sigma_xs_0+\frac{1}{\sqrt{6}}\sigma_ys_0\right)\nonumber\\
&&+2t_{\mathrm{SOC}}(-\sqrt{2}k_x+\sqrt{2}k_y)\sigma_zs_z.
\end{eqnarray}
The velocity operator is given by $v^x=t\sigma_xs_0/\sqrt{2}\hbar-t\sigma_ys_0/\sqrt{6}\hbar-2\sqrt{2}t_{\mathrm{SOC}}\sigma_zs_z/\hbar$ and $v^y=t\sigma_xs_0/\sqrt{2}\hbar-t\sigma_ys_0/\sqrt{6}\hbar+2\sqrt{2}t_{\mathrm{SOC}}\sigma_zs_z/\hbar$. Therefore, $\eta(\mathrm{M})=(t^2-12t^2_{\mathrm{SOC}})/(t^2+12t^2_{\mathrm{SOC}})$ and $\eta(\mathrm{M})$ can take the value of 1 when $t_{\mathrm{SOC}}=0$. Similarly, if we rotate the lattice anticlockwise by 15$^{\circ}$, $\eta(\mathrm{M'})=-(t^2-12t^2_{\mathrm{SOC}})/(t^2+12t^2_{\mathrm{SOC}})$ and $\eta(\mathrm{M'})$ can take the value of -1 when $t_{\mathrm{SOC}}=0$. 

When $t_{\mathrm{SOC}}=0$, the four-band model manifests as a superposition of two decoupled two-band models, enabling complete selectivity for linearly polarized light along specific orientations at M and M$'$. In the absence of SOC, the optical transitions at M selectively allow light polarized along $\hat{\boldsymbol{x}}+\sqrt{3}\hat{\boldsymbol{y}}$, while the optical transitions at M$'$ select light polarized along $\hat{\boldsymbol{x}}-\sqrt{3}\hat{\boldsymbol{y}}$. However, for $t_{\mathrm{SOC}}\neq0$, SOC hybridizes the four bands, causing the optical transitions at M and M$'$ to lose complete linear polarization selectivity.

\section{Calculation details}

Our prediction of material candidates is based on density functional theory (DFT) methods using the Perdew-Burke-Ernzerhof\cite{PBE} form of the generalized gradient approximation (GGA), as implemented in the Vienna ab initio simulation package (VASP)\cite{vasp}. The Gamma scheme $k$-point mesh size is set to $21\times21\times1$, and the plane-wave cutoff energy is set to 600 eV. The Hubbard $U$ is set to 3 eV for V~\cite{xu2025chemical}. The crystal structure is optimized until the forces on the ions are less than 0.001 eV/$\mathring{\mathrm{A}}$. Then the Wannier model is constructed using wannier90~\cite{wannier90}. The matrix representations of the velocity operator are calculated according to Ref.~\cite{wang_initio_2006}. Then the quantum metric tensors are calculated using the velocity matrices.

\normalem
\bibliography{reference}